\documentclass[aps,showpacs,twocolumn,prd,superscriptaddress,nofootinbib]{revtex4}

\usepackage{amsmath}
\usepackage{latexsym}
\usepackage{graphicx}
\usepackage[usenames]{color}
\usepackage{url}

\newcommand{\beq}{\begin{equation}}
\newcommand{\eeq}{\end{equation}}
\newcommand{\bea}{\begin{eqnarray}}
\newcommand{\eea}{\end{eqnarray}}
\newcommand{\ba}{\begin{array}}
\newcommand{\ea}{\end{array}}
\def\lsim{\raise 0.4ex\hbox{$<$}\kern -0.8em\lower 0.62ex\hbox{$\sim$}}
\def\gsim{\raise 0.4ex\hbox{$>$}\kern -0.8em\lower 0.62ex\hbox{$\sim$}}
\newcommand{\lm}{\ell m}
\newcommand{\lopt}{\mathcal{L}}
\newcommand{\mopt}{\mathcal{M}}
\newcommand{\lmopt}{\mathcal{L}\mathcal{M}}
\newcommand{\ie}{i.~e.~}

\def\leq{\,\raise 0.4ex\hbox{$<$}\kern -0.8em\lower 0.62ex\hbox{$-$}\,}
\def\geq{\,\raise 0.4ex\hbox{$>$}\kern -0.8em\lower 0.62ex\hbox{$-$}\,}
\def\pm{\,\raise 0.4ex\hbox{$+$}\kern -0.8em\lower 0.62ex\hbox{$-$}\,}

\newcommand{\MSun}{M_{\odot}}
\newcommand{\imagi}{i}

\begin{document}

\title{Observing mergers of non-spinning black-hole binaries}

\author{Sean T. McWilliams}
\affiliation{Gravitational Astrophysics Laboratory, NASA Goddard Space Flight Center, 8800 Greenbelt Rd., Greenbelt, MD 20771, USA}
\email{Sean.T.McWilliams@nasa.gov}
\author{Bernard J. Kelly}
\affiliation{CRESST and Gravitational Astrophysics Laboratory, NASA Goddard Space Flight Center, 8800 Greenbelt Rd., Greenbelt, MD 20771, USA}
\affiliation{Department of Physics, University of Maryland, Baltimore County, 1000 Hilltop Circle, Baltimore, MD 21250}
\author{John G. Baker}
\affiliation{Gravitational Astrophysics Laboratory, NASA Goddard Space Flight Center, 8800 Greenbelt Rd., Greenbelt, MD 20771, USA}

\date{\today}

\begin{abstract}

Advances in the field of numerical relativity now make it possible to calculate the final, 
most powerful merger phase of binary black-hole coalescence for generic binaries.  The state of the art has advanced
well beyond the equal-mass case into the unequal-mass and spinning regions of parameter space.
We present a study of the nonspinning portion of parameter space, primarily using
an analytic waveform
model tuned to available numerical data, with an emphasis on observational implications. 
We investigate the impact of varied mass ratio on merger signal-to-noise ratios (SNRs) for several detectors, and compare
our results with expectations from the test-mass limit.  We note a striking similarity of the waveform phasing
of the merger waveform across the available mass ratios.  Motivated by this, we calculate the match
between our 1:1 (equal mass) and 4:1 mass-ratio waveforms during the merger as a function of location on the source sky,
using a new formalism for the match that accounts for higher harmonics.  This is an indicator of the amount of degeneracy in mass ratio
for mergers of moderate-mass-ratio systems.

\end{abstract}

\pacs{
04.30.Db, 
04.80.Nn  
95.30.Sf, 
97.60.Lf  
}

\maketitle

\section{Introduction}
\label{sec:intro}

The merger of a black-hole binary will be one of the strongest sources
of gravitational waves, with a greater luminosity than the combined
electromagnetic luminosity from all the stars in the visible universe.
Ground-based detectors like LIGO, Virgo, and GEO, currently entering their second generation
of development, are sensitive to the mergers of stellar black holes, while
the space-based LISA will observe mergers of
massive and supermassive black holes.
It has long been expected that the final mergers of black-hole binaries would be significant
for interpreting gravitational wave measurements.  While a physically motivated model,
and corresponding template bank, would not be necessary for detection \cite{Flanagan:1997sx},
such a model would be the only avenue toward extracting all of the available information about the system
that is contained in the merger signal.
And since the merger is likely to constitute the majority
of the detectable signal for the next generation of ground-based detectors \cite{Baker:2006kr}, 
such a physically motivated model would be necessary
for gaining an understanding of the physical sources generating the detected signals.

In the absence of merger models, early investigations had to use information
from perturbative approximations to guess at the impact of mergers.  In \cite{Flanagan:1997sx}, the
Newtonian approximation for the gradual adiabatic inspiral of the holes, combined with the understanding
of the post-merger ringdown as the quasi-normal modes of a Kerr black hole, was used to guess at the
contribution of mergers to the signal detectability.  This guess was essentially validated by the observed
behavior of numerically simulated merger signals \cite{Baker:2006kr}.  However, while the power spectrum could be approximated,
the physics behind the power spectrum, the amplitude and phase evolution that would lead to that spectrum,
the accuracy with which the merger phase could be simulated or modeled, and the amount of information about the source
that could be extracted from detected signals were completely open questions.
Over the course of the last few years, the field of numerical relativity has provided
a means of studying the detailed structure of these merger signals for the first time.
Initially focusing only on the equal-mass, nonspinning case, several groups have since
explored both the nonspinning axis of parameter space as well as the vast expanse of spinning
parameter space \cite{Baker:2006kr,Hinder:2007qu,Damour:2007vq,Hannam:2007ik,Hannam:2007wf,Gonzalez:2008bi,Scheel:2008rj,Chu:2009md}.
The current availability of merger waveforms now makes it possible
to address the questions previously mentioned, through the measurement of
these signals.  

Because the merger is the dominant contributor to the overall signal power, particularly for ground-based
detectors where it provides the majority of the detectable signal, answering these questions is
a critical exercise actively being addressed by many groups.  Significant attention has been given
to the problem of modeling the signals with sufficient accuracy for detection with ground-based observations.
Recent work has begun addressing not only detection, but also estimating the source parameters
using ground-based \cite{Ajith:2009fz} and space-based detectors \cite{McWilliams_PhD, Babak:2008bu,
Thorpe:2008wh,McWilliams:2009bg}.
Much work has also gone into interpreting the
available merger waveforms, in an attempt to better understand them, both with regard to a physical
interpretation of the source \cite{Baker:2008mj} and with regard to 
understanding what drives the recoil of systems due to asymmetric radiation
\cite{Berti:2007nw,Schnittman:2007ij,Brugmann:2007zj}.

In this paper, we revisit the nonspinning subset of parameter space, with the goal of studying the
observational implications of nonspinning merger waveforms.  In Sec.~\ref{sec:method}, we briefly describe
the procedure for generating complete nonspinning waveforms.  In Sec.~\ref{sec:SNR}, we study the contribution
to the achievable SNR from the inclusion of mergers, and its variation with mass ratio.
In Sec.~\ref{sec:phase}, we study
more detailed comparisons of the nonspinning waveforms, including a novel implementation of the ``match''
statistic \cite{Owen:1995tm}.
In Sec.~\ref{sec:conc}, we discuss the general observational implications of nonspinning merger waveforms.
In the Appendix, we derive the formalism for the novel match implementation employed in Sec.~\ref{sec:phase}. 

\section{Methodology}
\label{sec:method}

The observable quantity being measured by gravitational wave interferometers, be they ground- or space-based, is the strain
on the spacetime, $h=\delta L/L$, or its derivatives.  We therefore require a model of the waveform, $h(t)$,
that we expect to measure from black-hole binaries.  The models we employ are predictions
for the \emph{emitted} strain, or the strain in the source frame.  
The \emph{detected} strain depends on the distance to the source,
the position on the detector's sky, and the detector's response.

In this work, we focus on 
Advanced LIGO for ground-based observation, and LISA for space-based observation.
For Advanced LIGO, we assume a constant response
as a function of frequency, which should be adequate for all but the lowest-mass cases.
For the detector noise, we
use the wide-band tuning \cite{DHSCom} typically associated with burst sources as was 
done in \cite{Baker:2006kr}, due to the
superior sensitivity at higher frequencies where the merger will occur for lower masses.
For LISA, we employ the effective noise floor from \cite{Larson:1999we, LISASenGen}, which includes both the
contributions from noise sources as well as the average response of the detector to
signals and instrumental noise, which is non-trivial for 
the higher frequencies in LISA's band.  For frequencies in the range 
$3\times 10^{-5}{\rm Hz} \leq f \leq 1\times 10^{-4}{\rm Hz}$, we employ a more conservative estimate 
of the acceleration noise as was done in \cite{Baker:2006kr}, instead assuming a steeper amplitude
spectral density that falls off as $f^{-3}$ \cite{MerkowitzCom}. 
Below $3\times 10^{-5}{\rm Hz}$, we assume the detector has no sensitivity.
We apply an overall factor of 3/20 to the LISA power spectral density as discussed in \cite{Berti:2005ys}. 

The remaining element is the model of the emitted waveform, which will depend on the intrinsic
parameters (i.~e.~the mass, mass ratio, spin vectors, and eccentricity), and which will vary over the sky of the source.
The emitted waveform can be conveniently represented by a harmonic mode decomposition.  If $h$ is the complex 
strain, then the mode decomposition is given by
\begin{equation}
h = \sum_{\ell=2}^{\infty} \sum_{m=-\ell}^{\ell} h_{\ell m}(t,R) \,^{-2}Y_{\ell m}(\theta,\phi) \, ,
\label{eqn:hlm_def}
\end{equation}
where $^{-2}Y_{\ell m}$ are the spherical harmonics of spin-weight (-2) \cite{Goldberg:1967}.
Being complex, $h$
contains both wave polarizations, defined by the relationship $h\equiv h_+ + \imagi h_{\times}$.
For an equal-mass system, $h(t,R,\theta,\phi)$ is dominated by the \emph{quadrupole},
the combination of $\ell=2$, $m=\pm2$ modes:
\begin{equation}
h_{\rm quad} = h_{2 2} \,^{-2}Y_{2 2}(\theta,\phi) + h_{2 -2} \,^{-2}Y_{2 -2}(\theta,\phi).
\label{eqn:hquad_def}
\end{equation}
Additionally, symmetry considerations for equal-mass nonspinning systems demand that 
$h_{2 2} = h_{2 -2}^*$; therefore we will often use $h_{2 2}$ as a proxy for the full quadrupole
waveform.

\begin{figure}
\includegraphics[scale=.25, angle=0]{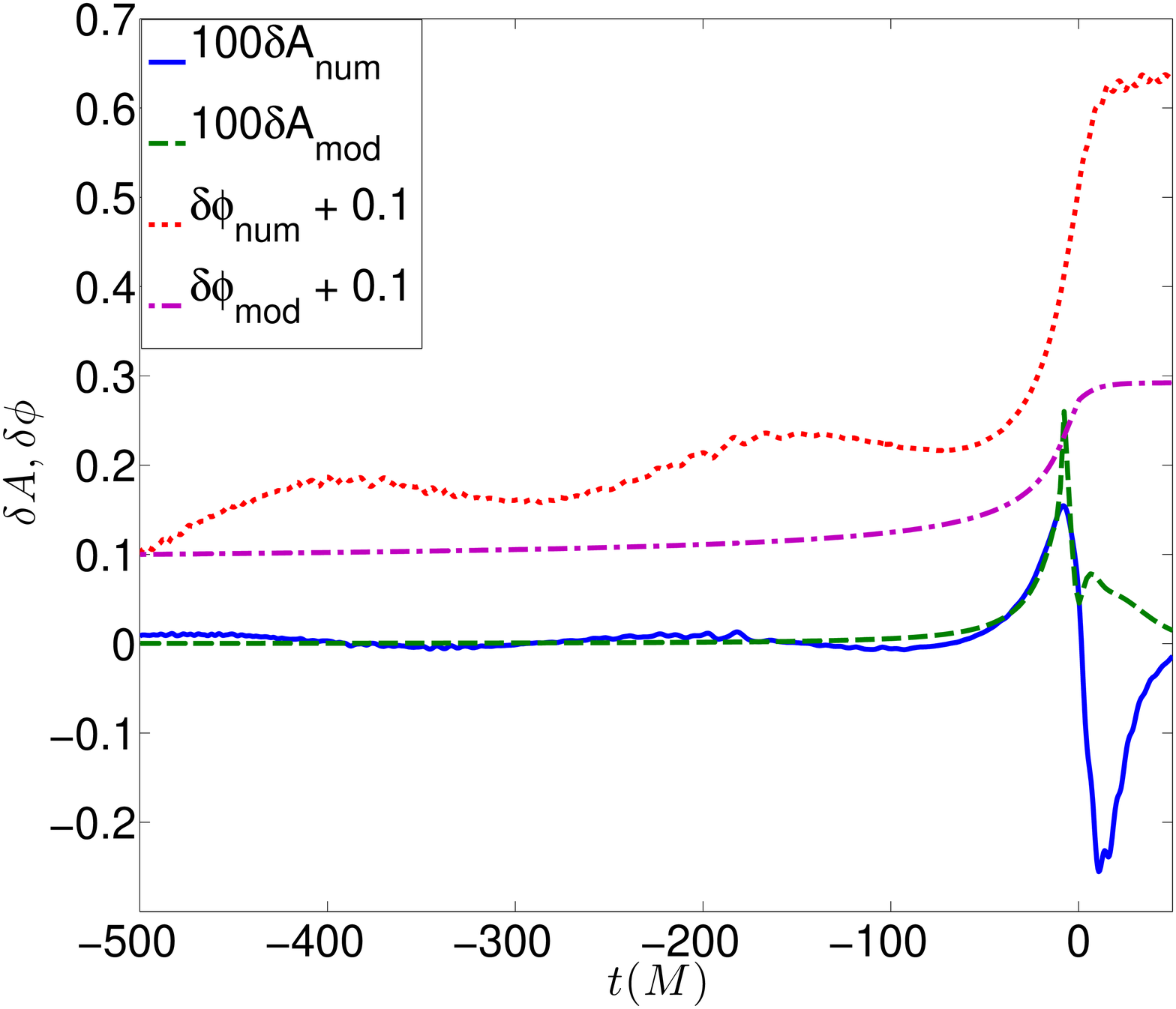}
\caption
{Comparison of amplitude and phase errors between the numerical data and the IRS-EOB model, for
the case of a 4:1 waveform. 
The inherent phase inaccuracy of the model is significantly
smaller than the numerical phase error, but the amplitude errors are comparable for the model and numerical data.}.
\label{fig:numvsmod}
\end{figure}

For the waveform comparisons presented in this work,
we use the model first presented in \cite{Baker:2008mj}, which has been validated by comparison with 
available data from numerical simulations for all harmonic components
through $\ell=4$. The model, referred to as the
IRS-EOB model, uses the effective-one-body (EOB) Hamiltonian formalism for the
inspiral \cite{Buonanno:2007pf}.  For the merger-ringdown, we employ a novel paradigm which we call the
implicit rotating source (IRS), wherein we apply a fit to a physically-motivated
functional form for the phasing (Eq.~9
in \cite{Baker:2008mj}), and, for the amplitude, a model for the flux
constrained to be consistent with the inspiral flux through
3.5 post-Newtonian (PN) order and to vanish as it approaches the ringdown frequency (referred
to as ``Model 2'' and given by Eq.~19 in \cite{Baker:2008mj}).  In Figure \ref{fig:numvsmod},
we compare the intrinsic error in phase and amplitude for the model and for our numerical waveforms,
using 4:1 as a representative case.  For brevity, we will refer to the unequal-mass runs as ratios, 
i.~e.~ the $q\equiv M_1/M_2=1/4$ run will be the 4:1 run, and the ratio notation will only be used in this context.
We shift all waveforms to peak in amplitude at $t=0$, and to
agree in phase at $t=-500M$.  We use the difference between our two highest resolutions as an indicator
of our numerical error, and we assume errors in the model due to different model parameters are independent,
and calculate a phase error 
$\delta\phi=\sqrt{\sum_i\left(\frac{\partial\phi}{\partial\lambda_i}\delta\lambda_i\right)^2}$, where $\lambda_i$
is simply the $\lambda$ parameter as used in \cite{Buonanno:2007pf} for the EOB inspiral, and 
$\lambda_i=\{\kappa,\,b,\,t_o,\,\dot{\Omega}_o,\,\Omega_f\}$ for the IRS merger, using the notation
in \cite{Baker:2008mj}.  For $\delta\lambda_i$, we use the values in Table II of \cite{Baker:2008mj}.

\begin{figure*}
\begin{center}
\includegraphics*[scale=.40, angle=0]{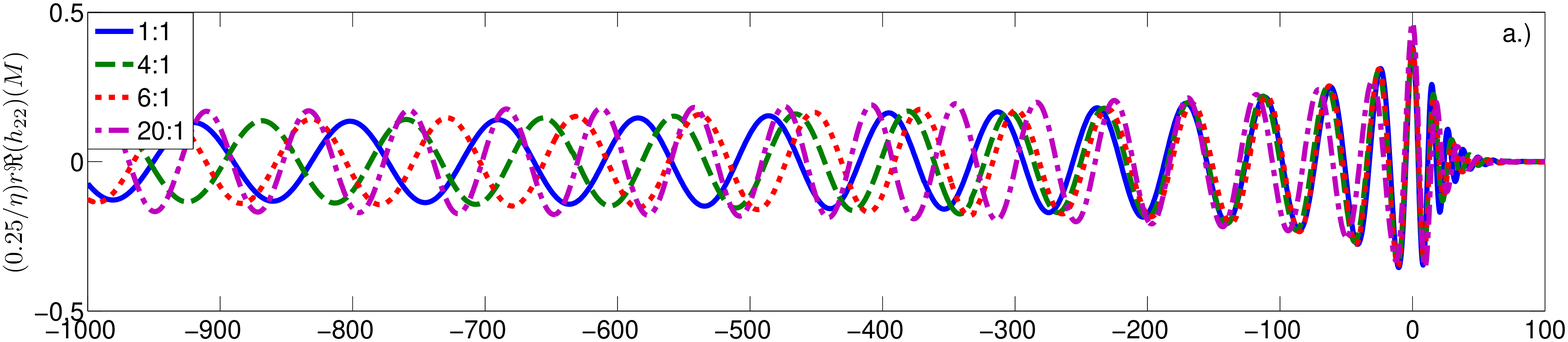}
\includegraphics*[scale=.40, angle=0]{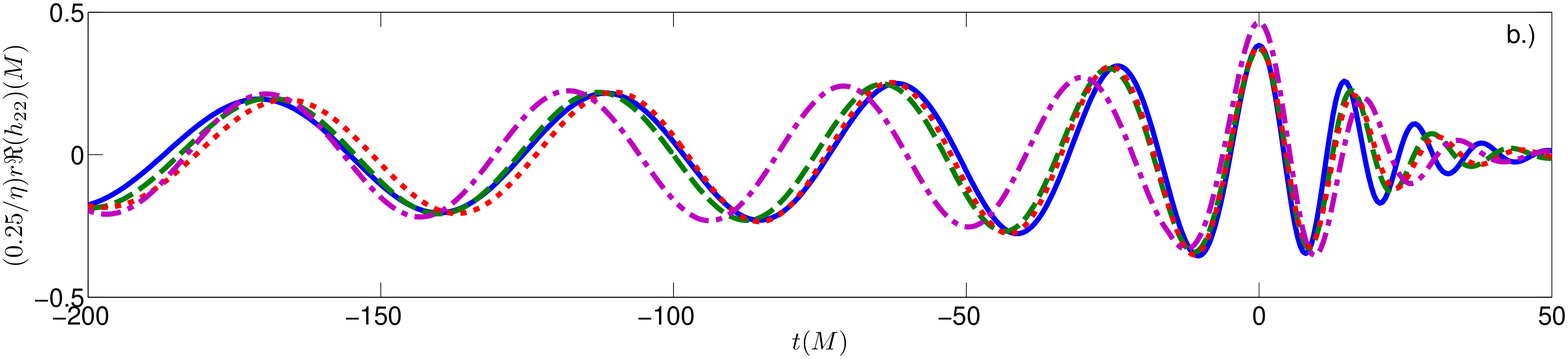}
\caption
{In the top panel (a.), we show quadrupole waveforms generated by using the model presented in 
\cite{Baker:2008mj} for mass ratios of 1:1 (equal mass), 4:1, 6:1, and 20:1.
When the waveforms are aligned in time based on their peak amplitudes, and aligned in phase to agree 
at said time, there is
significant overlap of the waveforms
for the 1:1, 4:1, and 6:1 cases over the final $\sim 5$ cycles leading up to merger, which is shown more clearly
in the bottom panel (b.).}
\label{fig:waves}
\end{center}
\end{figure*}

Figure~\ref{fig:waves} shows the quadrupole radiation for four mass ratios
-- 1:1, 4:1, 6:1, and 20:1 -- generated using the IRS-EOB model.
We note that in using the 20:1 mass ratio, we have extrapolated
to mass ratios that cannot, as yet, be validated by simulation.
The amplitudes for all the runs in Figure \ref{fig:waves} 
have been rescaled to better agree with the equal-mass amplitude, 
using the leading-order Newtonian scaling.  This also emphasizes the phasing agreement that begins in the 
late inspiral
and continues through the merger waveform, which was discussed in \cite{Baker:2008mj} and will be the topic of 
further discussion in a later section.

Since many investigations
relating to both LIGO and LISA have focused on detectability, rather than characterizing the signal, a model
of a quadrupole-only signal has been adequate within their margin of error.
However, higher harmonics can be more 
significant for calculations such as determining template fidelity with the match, 
or any attempt to extract source parameters from the signal,
as such investigations depend sensitively on the fine detail of the phase evolution.  We will investigate
the impact of higher harmonics in this context, and we will also include
higher harmonics in our calculations of SNR,
although the SNR contribution is essentially negligible for all cases investigated here with the possible
exception of 20:1.

\section{SNR and power scaling}
\label{sec:SNR}

SNR is the most useful statistic for assessing the detectability
of a given signal with a particular detector.
The SNR, which we denote as $\rho$, is given by
\beq
\label{eqn:snr_def}
  \rho^2 = \langle h | h \rangle,
\eeq
where ``$\langle\cdot|\cdot\rangle$'' denotes a noise-weighted inner product, given by
\beq
\langle h_{1}|h_{2}\rangle \equiv 2\int\limits_{0}^{\infty} df\, \frac{\tilde{h}_{1}^{*}\tilde{h}_{2} + \tilde{h}_{1}\tilde{h}_{2}^{*}}{S_{n}} \, , 
\label{eqn:dotprod}
\eeq
and $S_n(f)$ is the power spectral density of the detector noise discussed in the previous section.
The sky-averaged SNR is given by
\beq\label{eqn:snr}
  \langle\rho^2\rangle = \int\limits_{0}^{\infty} d(\ln f)\,\left(\frac{h_{\rm char}(f)}{h_{n}(f)}\right)^2.
\eeq
Here, $h_{\rm char}(f)\equiv 2\,f|\tilde{h}_{\rm opt}(f)|$ is the
characteristic signal strain, and $h_{n}(f)\equiv \sqrt{5}h_{\rm rms}(f) =
\sqrt{5fS_n(f)}$ is the root-mean-square of the detector noise fluctuations
multiplied by $\sqrt{5}$ for sky-averaging.
$\tilde{h}_{\rm opt}(f)$ is the Fourier transform of the optimally-oriented signal strain
\cite{Flanagan:1997sx}.  For quadrupole-only cases, the orientation-averaged 
signal strain is trivially calculated from the optimally-oriented
strain by dividing by $\sqrt{5}$.

Before the advent of merger waveforms from numerical relativity, expectations
about the power scaling of the merger waveforms, and thus the achievable SNRs,
were formed by using the test-mass limit as a surrogate, while the scaling of
the inspiral power can be approximated by PN expansions in the weak-field
limit.  Specifically, we know that the SNR from the inspiral scales as
$\sqrt{\eta}$ to leading order,
where $\eta \equiv M_1 M_2/M^2$ is the \emph{symmetric mass ratio} of the binary.
It was further
assumed in \cite{Flanagan:1997sx}, based on the prediction for total radiated
energy in the test-mass limit \cite{Detweiler79}, that the merger SNR scales
as $\eta$.
We note, however, that in \cite{Baker:2008mj}, the 
peak (2,2)-mode energy flux was best fit by the function
\beq
\dot{E}_{22} = 4.40 \times 10^{-3} \eta^2 + 5.43 \times 10^{-2} \eta^4\,,
\label{eqn:Edot22_quartic_fit}
\eeq
with $\dot{E}_{22}\propto |\dot{h}_{22}|^2 \propto \rho^2$.
The significant improvement in performance of the $\eta^2 + \eta^4$ fit,
compared to a strictly $\eta^2$ fit, may indicate 
that differences between the physics of
the merger for comparable masses and the test-mass plunge are being measured.
The absence of a well-defined innermost stable
circular orbit (ISCO) in the equal-mass case \cite{Damour:2000we,Cook:2001wi}, 
compared to the obvious ISCO for sufficiently small mass ratios, further
supports this picture. Certainly as the masses differentiate more, the
test-mass analogy bears out more.  For the equal-mass case,
Eq.~\eqref{eqn:Edot22_quartic_fit} indicates that the $\eta^2$ and $\eta^4$ terms
contribute roughly equally, but 
the $\eta^4$ term obviously becomes less
important for ever-smaller mass ratios.

\begin{figure*}
\begin{center}
\includegraphics*[scale=.26, angle=0]{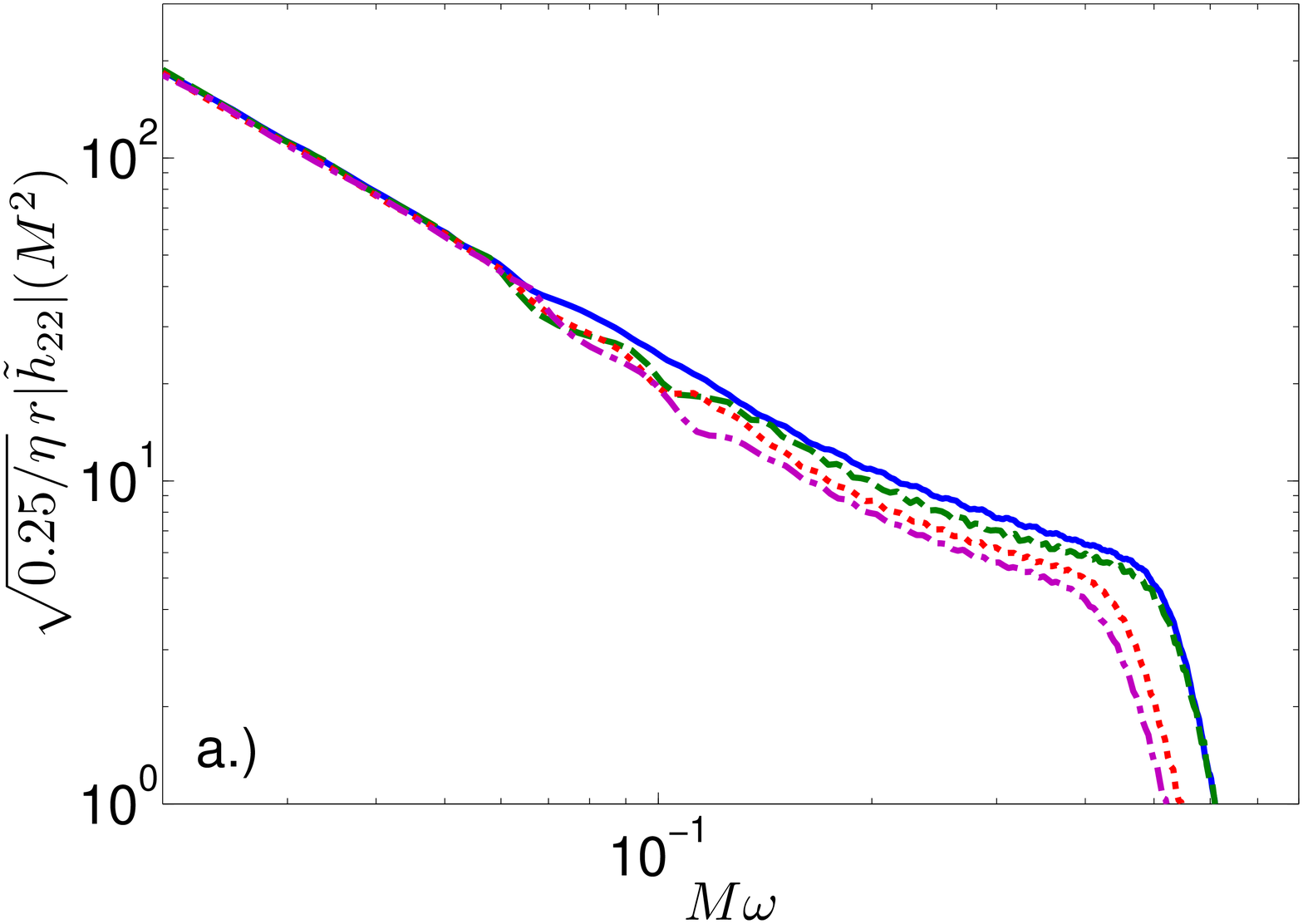}
\includegraphics*[scale=.26, angle=0]{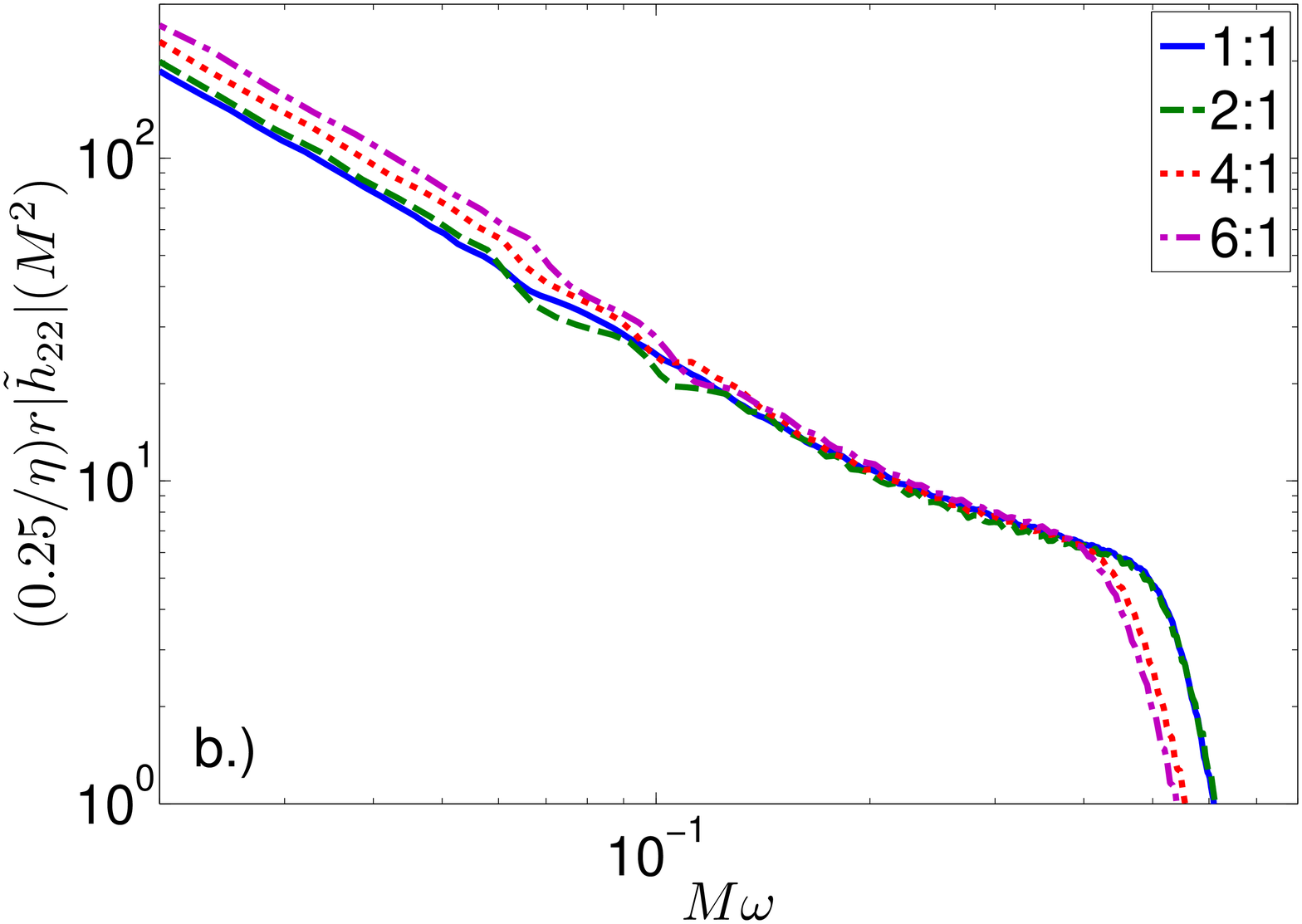}
\caption{Scaling of the Fourier amplitude of hybrid (PN inspiral/numerical merger) waveforms for different mass ratios by $\sqrt{\eta}$ in the left panel (a.), 
and by $\eta$ in the right panel (b.), which
appears to be an excellent approximation.}
\label{fig:hfeta}
\end{center}
\end{figure*}

The different scalings with $\eta$ are illustrated in Figure \ref{fig:hfeta}, where
we plot the Fourier transform of hybrid waveforms, constructed analogously to
\cite{Baker:2006kr} by tying a PN inspiral to our numerical data at a point where
they reach equal accuracy.  We do this in part to emphasize that the change in scaling of the
merger signal is not an artifact of the IRS-EOB model, but is apparent in our raw
numerical waveforms.  The left panel demonstrates the $\sqrt{\eta}$ scaling
of the inspiral ($M\omega \lesssim 0.08$) for the 1:1, 2:1, 4:1, and 6:1
quadrupolar waveforms, and the right panel shows that the merger scaling
is well approximated by a linear dependence on $\eta$ for
the merger ($M\omega \gtrsim 0.08$).  The deviation of the peak $\dot{E}_{22}$
from a simple quadratic-in-$\eta$ scaling appears to be due primarily to differences in the frequency
of the peak, since the signals closely follow a linear-in-$\eta$ scaling when evaluated at the same frequency
prior to the peak.

For equal masses and moderate mass ratios, $h(t,R,\theta,\phi)\approx h_{\rm quad}(t,R)$,
so that averaging the SNR over the 
binary's orientation is trivial.  This may be sufficient for ground-based detectors if they are
primarily detecting stellar-mass black holes
($M \leq 100 \MSun$) due to the limited available mass range.  If intermediate-mass black holes
($100 \MSun \leq M \leq 10^4 \MSun$) exist, then smaller mass ratios ($q \ll 1$) may occur,
and higher-order
modes will contain progressively more power relative to the quadrupole as $q \rightarrow 0$. 
Rather than ignoring higher-order modes or introducing complexity by averaging over them,
we instead focus on the optimal orientation of the binary, and only average over the sky of the detector.

\begin{figure*}
\begin{center}
\includegraphics*[scale=.24, angle=0]{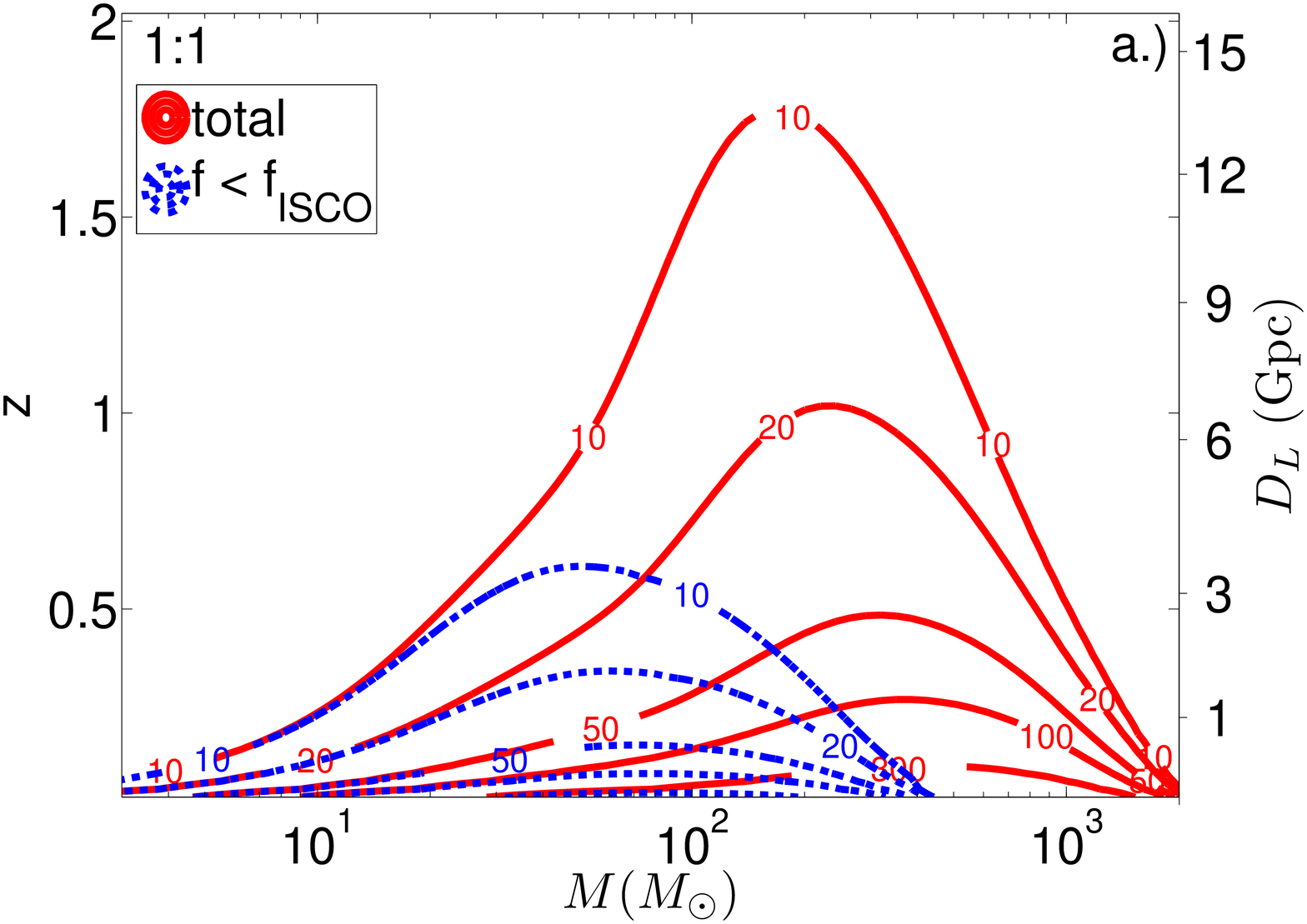}
\includegraphics*[scale=.24, angle=0]{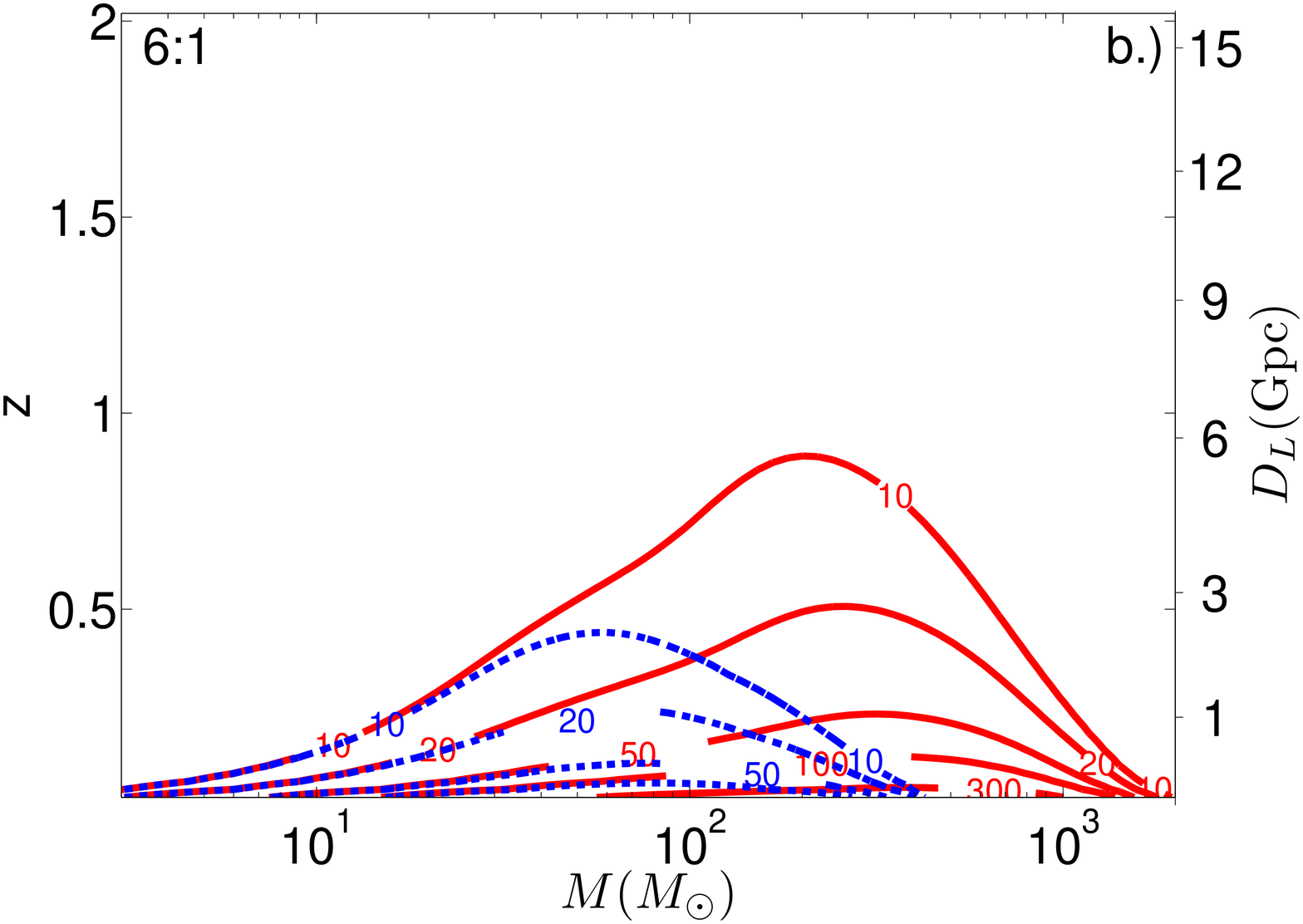}
\caption
{SNR contours for Advanced LIGO with $q = 1$ (a.) and $q = 1/6$ (b.).  Note, when comparing the two panels, 
that the masses are total masses, which determine the overall
waveform amplitude.  The solid lines correspond
to the SNR calculated from the full waveform, including the merger, while the dotted lines correspond to the
SNR contribution from the portion of the signal with frequency lower than the Schwarzschild ISCO frequency.} 
\label{fig:ADVLctr}
\end{center}
\end{figure*}

\begin{figure*}
\begin{center}
\includegraphics*[scale=.24, angle=0]{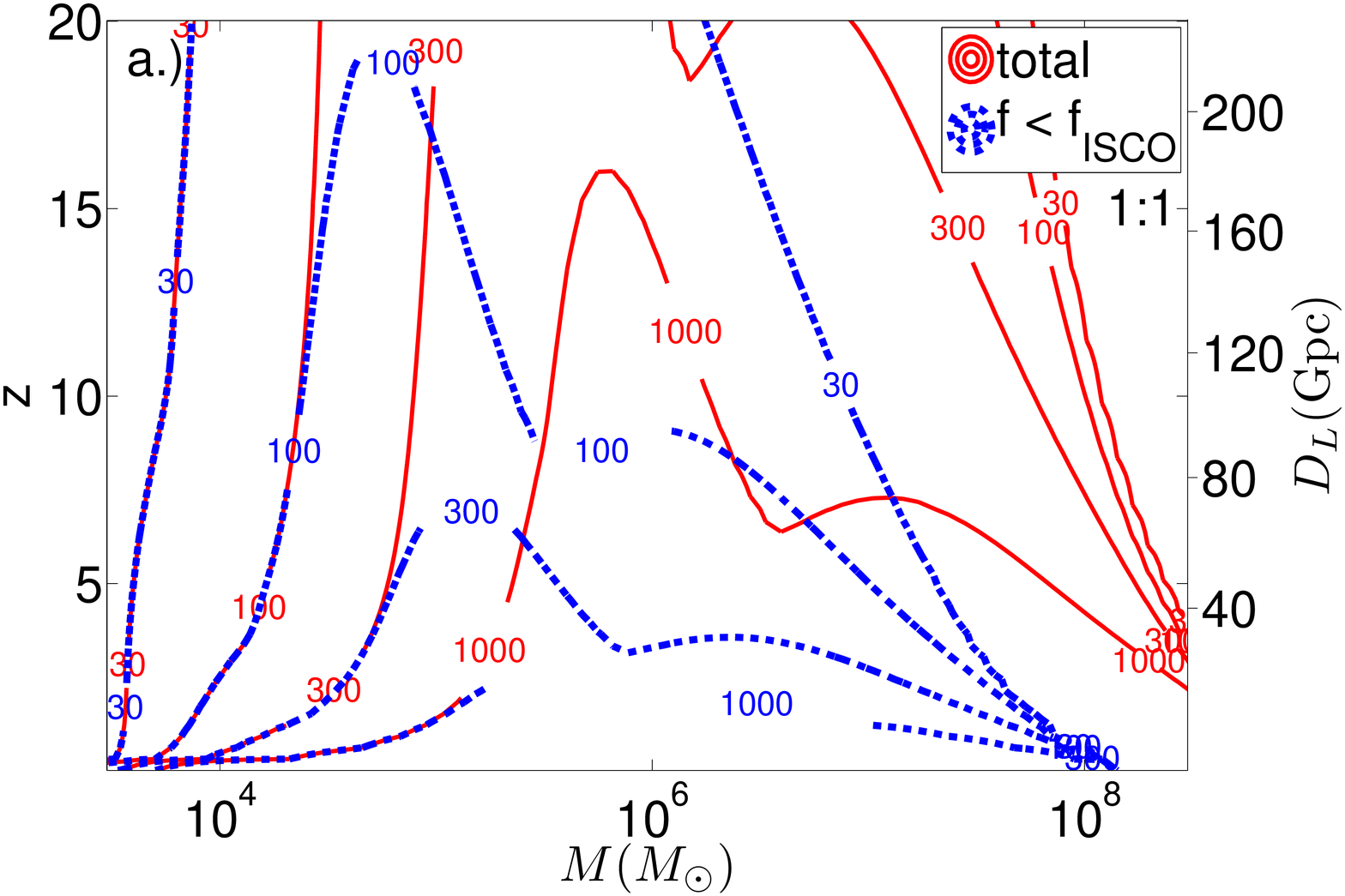}
\includegraphics*[scale=.24, angle=0]{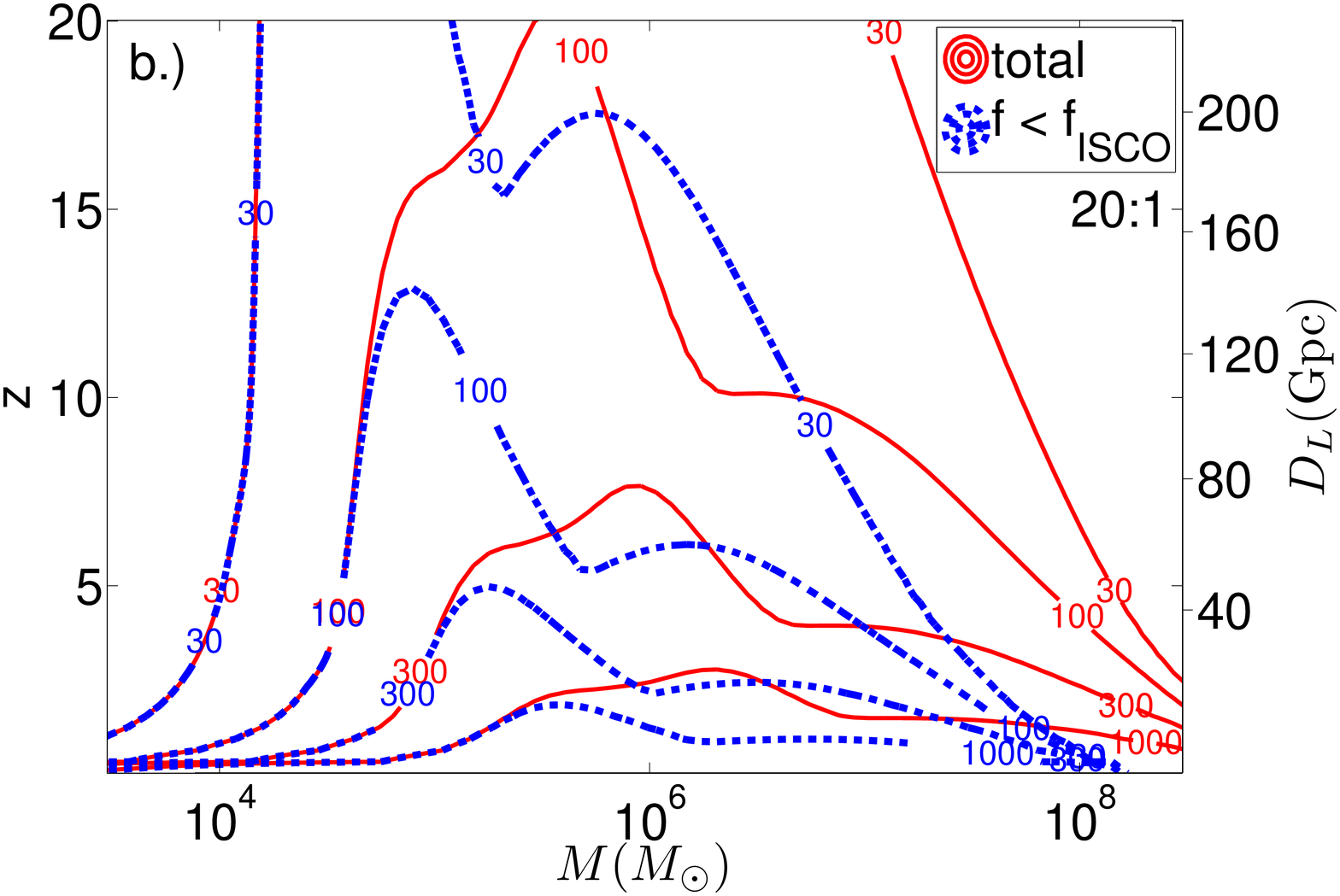}
\caption
{SNR contours for LISA with $q = 1$ (a.) and $q = 1/20$ (b.).  The solid lines again correspond
 to the full waveform SNR, while the dotted lines correspond to the
SNR contribution from frequencies lower than the Schwarzschild ISCO frequency.  While the observable range of high-SNR mergers is 
reduced by a factor of several at 20:1 from what was seen in the 1:1 case, sources are still easily detectable 
to large redshift over a similarly broad mass range.}
\label{fig:LISActr}
\end{center}
\end{figure*}

To demonstrate the decrease in SNR with a significant deviation from equal mass, we show the 1:1 and 6:1
cases for Advanced LIGO and the 1:1 and 20:1 cases for LISA in Figures \ref{fig:ADVLctr} and
\ref{fig:LISActr}, respectively. The panels show contour lines for both mass ratios, with one set of
lines corresponding to the SNR accumulated before the corresponding Schwarzschild ISCO, and the other
corresponding to the full signal. 
As described earlier, the SNR decreases as $\eta$ deviates from $0.25$, with the inspiral SNR scaling
as $\sqrt{\eta}$ and the merger SNR scaling as $\eta$.  
We therefore expect the SNRs to scale roughly as $\sqrt{\eta}$ for the lowest
masses where the inspiral matters most, and as $\eta$ (potentially with terms of higher power in $\eta$ as well) 
for higher masses where the merger contributes the
majority of SNR.  

In both mass-ratio cases for each detector,  the late inspiral-merger 
phase constitutes the majority of the SNR in high-SNR events.  
The merger contributes significantly for masses
$M \gtrsim 30 \MSun$ for Advanced LIGO, or $(1+z)M \gtrsim 10^5 \MSun$
for LISA. As previously reported for equal-mass mergers
\cite{Flanagan:1997sx,Baker:2006kr}, the merger contribution to the signal
tends to dominate strongly for these larger-mass systems.
For the unequal-mass cases the merger plays a dominant role over a similar 
range of masses, though the level to which the merger dominates
the overall SNR is significantly diminished for very unequal masses (right panels of Figures \ref{fig:ADVLctr}
and \ref{fig:LISActr}) compared to the equal-mass case (left panels; see also the figures presented in
\cite{Flanagan:1997sx,Baker:2006kr}). 
In some ways the equal-mass case is exceptional, rather than representative.
For observations of IMBH mergers ($M \gtrsim 100 \MSun$) with Advanced LIGO, however,
the merger always dominates, as the relatively sharp wall in low-frequency 
sensitivity effectively wipes out the inspiral contribution.

\section{Waveform Comparison}
\label{sec:phase}

Numerical relativity now provides
a reasonably clear picture of the late stages of merger, specifically in the form of the waveforms.
While considerable progress has
been made in understanding how to detect inspiral signals and characterize
how they depend on system parameters, including mass ratio and spin, 
there is little similar work addressing observations over the signal-space of mergers.  
Most observational work so far has considered these effectively as unmodeled sources.  In \cite{Baker:2008mj},
we examined the relationships between the merger waveforms and the physical motion of the source, 
emphasizing simple common features in order to form a general characterization of nonspinning mergers. 
These features have observational consequences as well.
In particular, we noted general similarity in the late-time portions of dynamics and waveforms,
which we now revisit.  

\begin{figure*}
\begin{center}
\includegraphics*[scale=.40, angle=0]{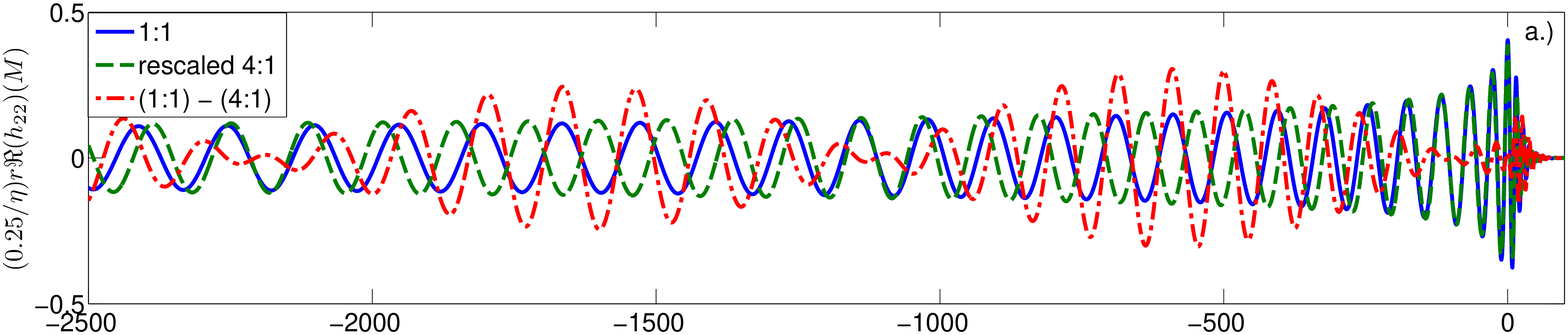}
\includegraphics*[scale=.40, angle=0]{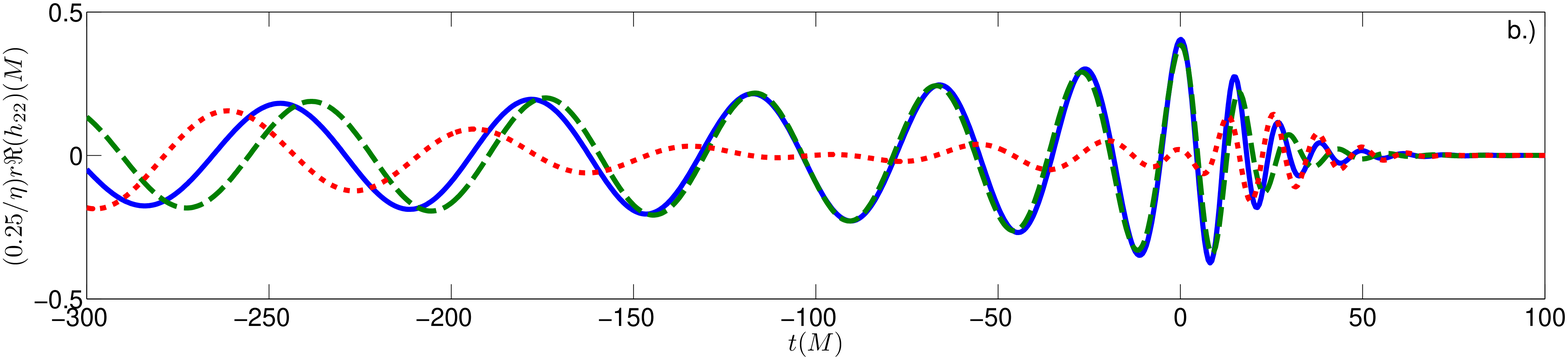}
\caption{Differences, $\delta h$, in 1:1 and rescaled 4:1 mass ratios.  The inspiral waveform evolves out of phase 
on a timescale shown clearly by the beats in $\delta h$ (a.).  The phase alignment persists for a slightly
longer time during the merger (b.).
}
\label{fig:hdiff}
\end{center}
\end{figure*}

\begin{figure}
\begin{center}
\includegraphics[scale=.25, angle=0]{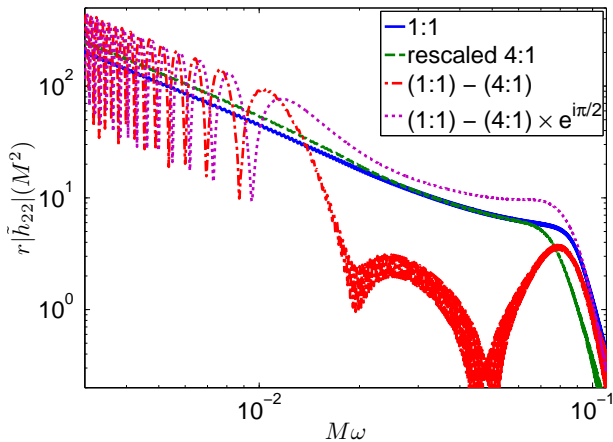}
\caption{The Fourier amplitude of the waveform difference in Figure \ref{fig:hdiff}, both with the phase shift shown there, as well as shifted by $\pi/2$ to illustrate the dependence of the power at high frequencies on the phase alignment.}
\label{fig:hdifffreq}
\end{center}
\end{figure}

In Figure \ref{fig:hdiff} we plot model waveforms for the 1:1 case, the 4:1 case with its
amplitude rescaled with the leading-order $\eta$ dependence, and the difference $\delta h=h_1-h_2$ between these two waveforms.
By doing so, we see that the apparent phase agreement shown in \cite{Baker:2008mj}
is, not surprisingly, partially an artifact of aligning all the phases at the peak strain amplitude, and thereby enforcing a node
in $\delta h$ at that time.  However, the merger is unique in the suppression
of the final beat prior to ringdown in $\delta h$, and enhancement of that beat if a $\pi/2$ phase shift is applied.
This can be seen in the time series of Figure \ref{fig:hdiff}, but is most evident in the Fourier representation
of Figure \ref{fig:hdifffreq}.  This extended frequency agreement also provides a simple explanation
for the $\sqrt{\eta}$ difference in scaling between the inspiral and merger.  The amplitude
scales linearly with $\eta$ to a good approximation for both the inspiral and merger, but the time interval spent
within a given frequency bin scales as $\eta^{-3/8}$ to leading order for the inspiral and is nearly constant for the merger
for moderate mass ratios.  This results in a relative $\eta^{5/8}$ amplitude scaling of $\tilde{h}$, and therefore SNR,
between the inspiral and merger.

\begin{figure*}
\begin{center}
\includegraphics*[scale=.25, angle=0]{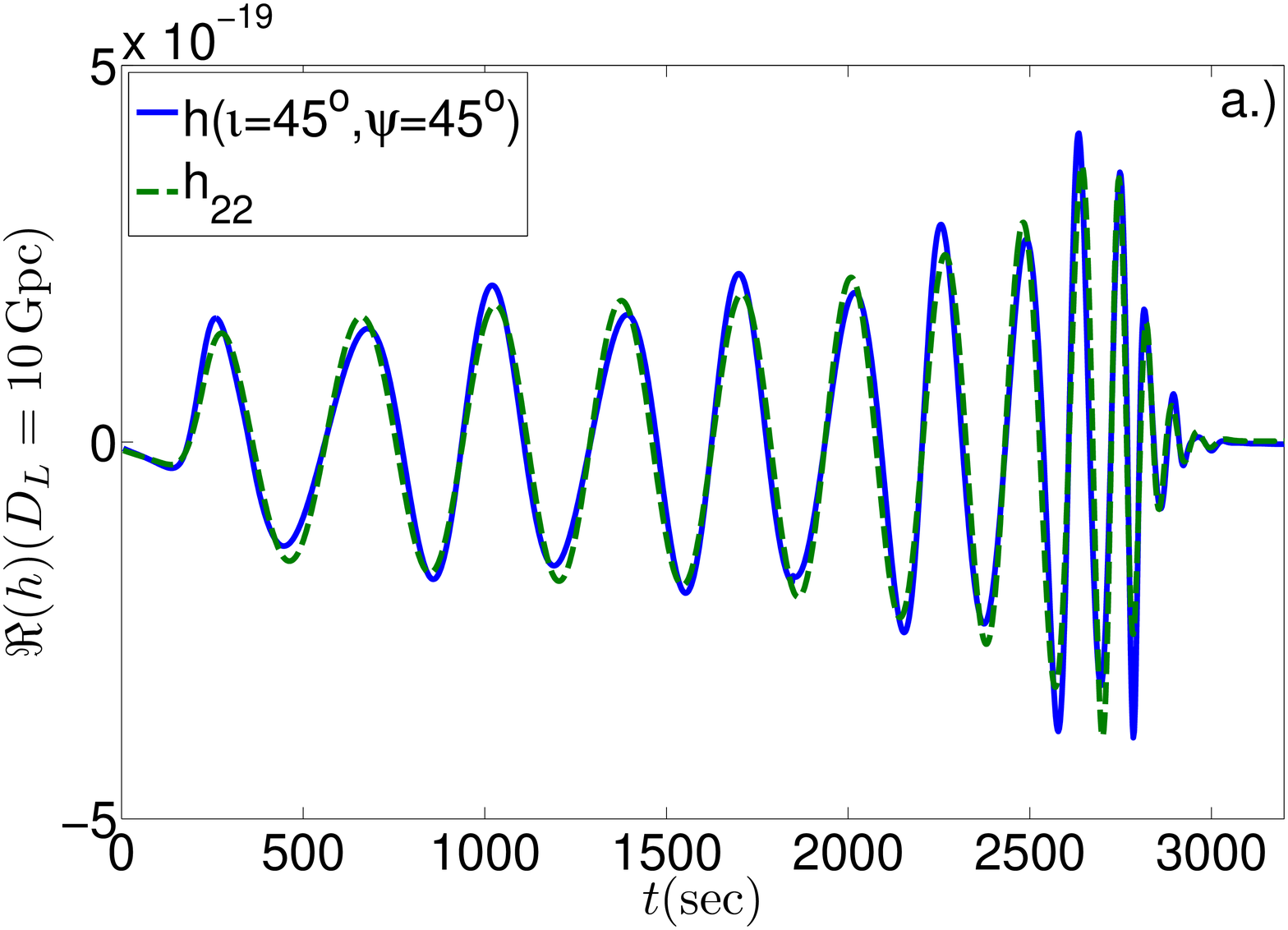}
\includegraphics*[scale=.25, angle=0]{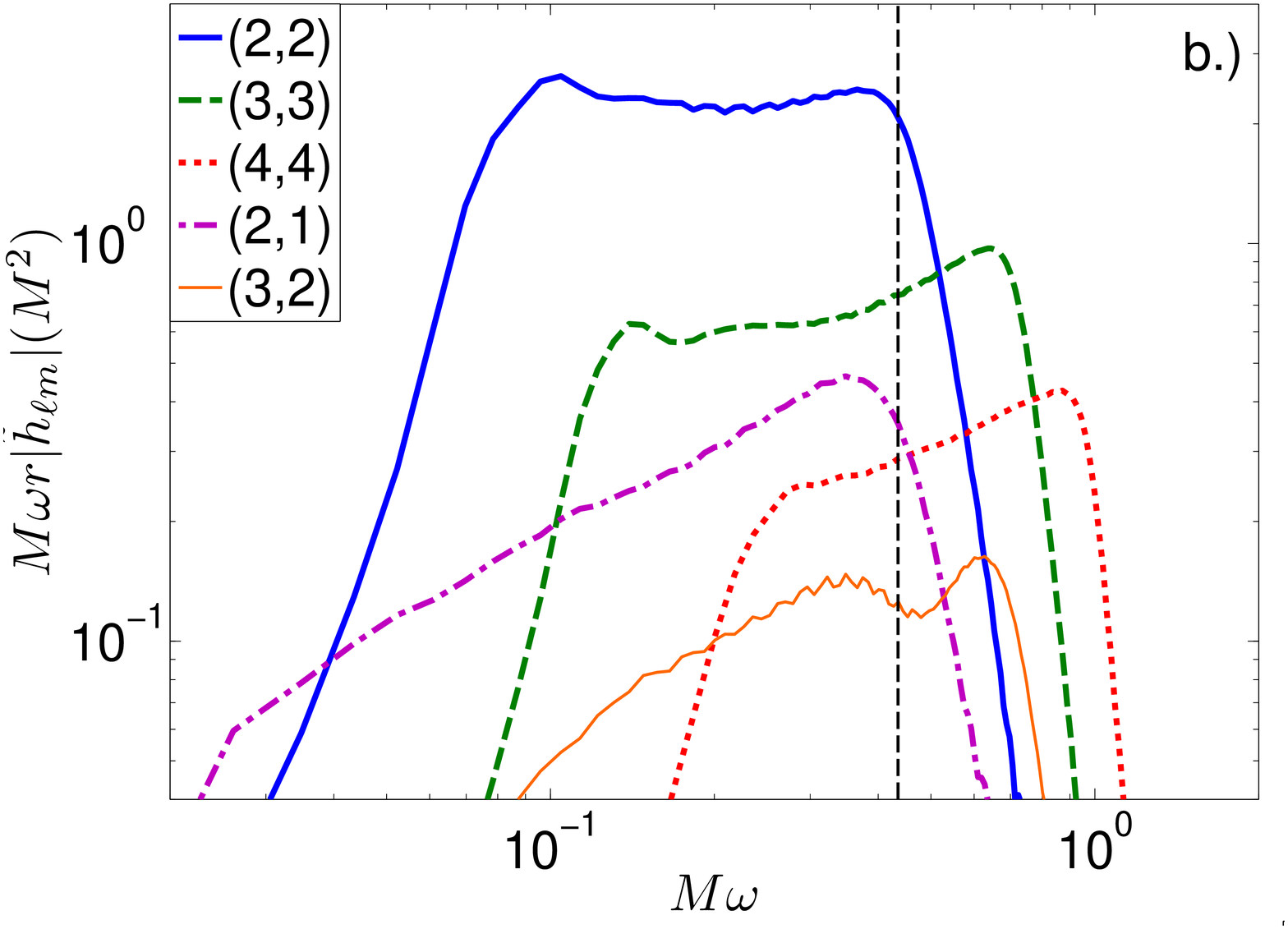}
\caption
{In the left panel (a.), we show a time series representation of a 6:1 mass-ratio $10^6 \MSun$ black-hole binary at a
distance of $10$ Gpc, using the waveforms from our numerical simulation.
This is a typical example of the $(2,\pm2)$ modes constituting the vast
majority of the overall power content of the waveform.  This is further demonstrated in the Fourier-series
representation shown in the right panel (b.), where the $(2,\pm2)$ modes dominate until well into the ringdown,
the onset of which is approximately indicated by the dashed vertical line.}
\label{fig:h6to1}
\end{center}
\end{figure*}

While SNR is certainly the most relevant statistic for detection purposes, it tells us
little about the details of the waveform, in particular the evolution of the phase,
which may be critical when answering questions regarding signal characterization.
The ``match'' \cite{Owen:1995tm} is a useful statistic for more detailed waveform comparisons,
as it is sensitive to small differences in waveform phase.  For any two waveforms, $h_1$
and $h_2$, the match ${\cal M}$ is defined using the noise-weighted inner product \eqref{eqn:dotprod}:
\begin{equation}
{\cal M} = \frac{\langle h_{1}|h_{2}\rangle}{\sqrt{\langle h_{1}|h_{1}\rangle\langle h_{2}|h_{2}\rangle}} \, .
\label{eqn:matchdef}
\end{equation}
The match can be viewed as the fraction of the matched-filter SNR that is recovered by using
$h_{2}$ as a filter to search for $h_{1}$, rather than using $h_{1}$ itself
(the optimal filter).  The left panel of Figure \ref{fig:h6to1} shows a typical comparison for the 6:1 case,
which should have the strongest higher harmonics among the numerical simulations studied here.
Also, we show a frequency-based comparison for the same case in the right panel of Figure \ref{fig:h6to1},
where the $(2,\pm 2)$ modes can be seen to dominate the signal power until well into ringdown
(indicated here by a vertical dashed line).  Nonetheless, it is still possible that a sub-dominant
mode may modulate the signal to a sufficient degree to significantly
diminish the recoverable SNR, or to impact the template member that has the highest likelihood,
if only the dominant mode is used as a filter.  We will therefore
develop an appropriate formalism for including all modes
analytically in a calculation of the match.

\begin{figure*}
\begin{center}
\includegraphics*[scale=.25, angle=0]{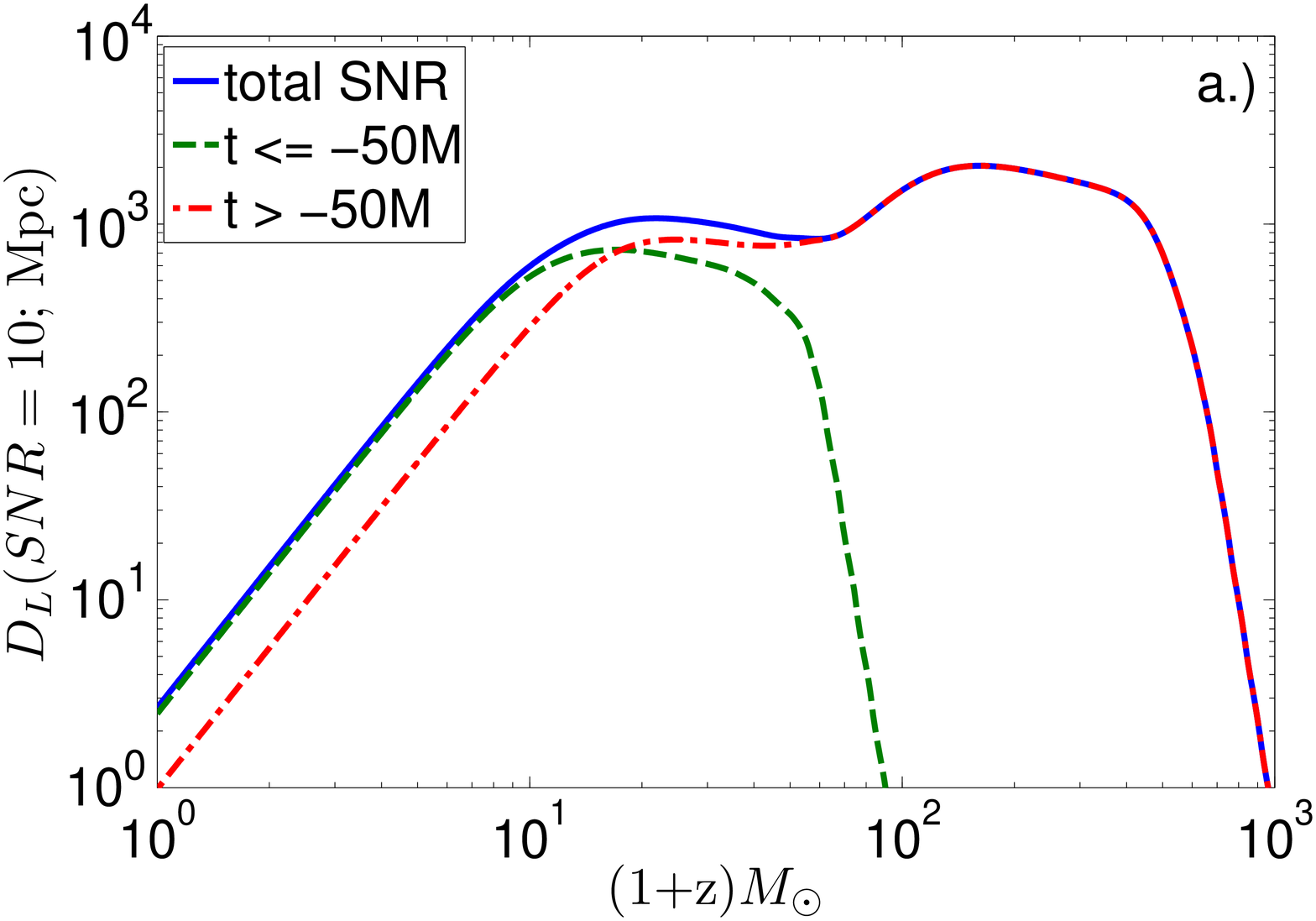}
\includegraphics*[scale=.25, angle=0]{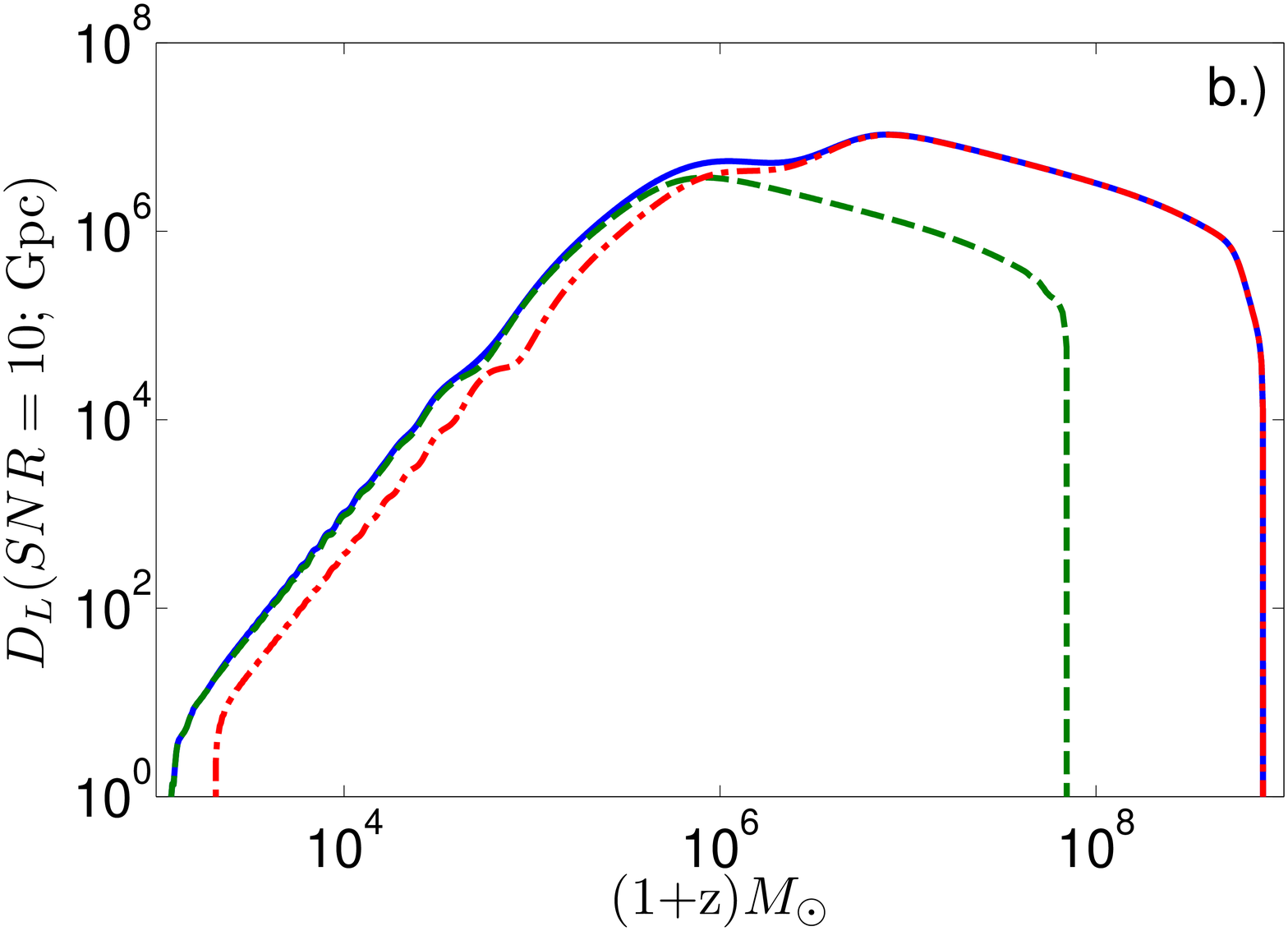}
\caption{Luminosity distance where the difference between the 1:1 and rescaled 4:1 mass ratio waveforms is detectable with an SNR of 10 for Advanced LIGO (a.) and LISA (b.), also referred to as the distance horizon.
This can be interpreted as being the maximum distance at which we can distinguish these two sources with each interferometer.}
\label{fig:LIGOhdiff}
\end{center}
\end{figure*}

We can further calculate the SNR of the difference in waveforms, $\delta h$ (see Figure \ref{fig:LIGOhdiff}),
which is essentially a measurement of our ability to distinguish two waveforms from each other.
This simple statistic is related to the ``mismatch'', $1-{\cal M}$, as well as the SNR, $\rho$ (see also \cite{Lindblom:2008cm}):
\bea
\langle \delta h | \delta h \rangle &\equiv& \langle h_1-h_2|h_1-h_2\rangle \nonumber \\
                                  &=&\langle h_1|h_1\rangle+\langle h_2|h_2\rangle-2\langle h_1|h_2\rangle  \nonumber \\
                                  &=&\left(|h_1|-|h_2|\right)^2 +2|h_1||h_2|\left(1-\frac{\langle h_1|h_2\rangle}{|h_1||h_2|}\right) \nonumber \\
                                  &\approx& 2\rho^2(1-{\cal M}),
\label{eqn:FFdiff}
\eea
where $|h_1|\equiv \sqrt{\langle h_1|h_1\rangle}$, and the final approximation comes from
assuming that the SNRs of $h_1$ and $h_2$ are approximately equal.
We note that the curves ``1:1'', ``rescaled 4:1'' and ``1:1 - 4:1'' in
Figure \ref{fig:hdifffreq}, are simply the integrands of 
$|h_1|$, $|h_2|$ and $\sqrt{\langle \delta h | \delta h \rangle}$,
respectively, without noise-weighting. The latter is roughly an order of magnitude smaller than
$|h_1|$ during the entire merger phase. In this case we can see from Eq.~\eqref{eqn:FFdiff}
that the match among
moderate-mass-ratio mergers is likely to be quite high for ground-based interferometers.
We can therefore expect that, for instance,
a small subset of merger waveforms would be capable of sufficiently covering a large range of
nonspinning parameter space for detection purposes,
but the apparent mass-ratio degeneracy in the merger will have a negative impact on parameter-estimation efforts.

The SNR has a trivial inverse proportionality with luminosity
distance, so swapping SNR and luminosity distance can give you, for instance, the distance horizon at which the difference between waveforms
can be detected, by setting the SNR at some fixed threshold.  We do so in Figure \ref{fig:LIGOhdiff}, using the same $\delta h$ as above
as an example.  We use a fixed SNR of $\rho = 10$ as the threshold of detectability.  The values in Figure \ref{fig:LIGOhdiff} have an interesting
implication in light of Eq.~\eqref{eqn:FFdiff}, in that for sources farther than the distance horizon, we cannot distinguish between
a 1:1 waveform at $D_L$, or a waveform with mass ratio $\eta$ at a distance $(\eta/0.25)D_L$. 

\begin{figure*}
\begin{center}
\includegraphics*[trim = 25mm 0mm 25mm 0mm, clip, scale=.40, angle=0]{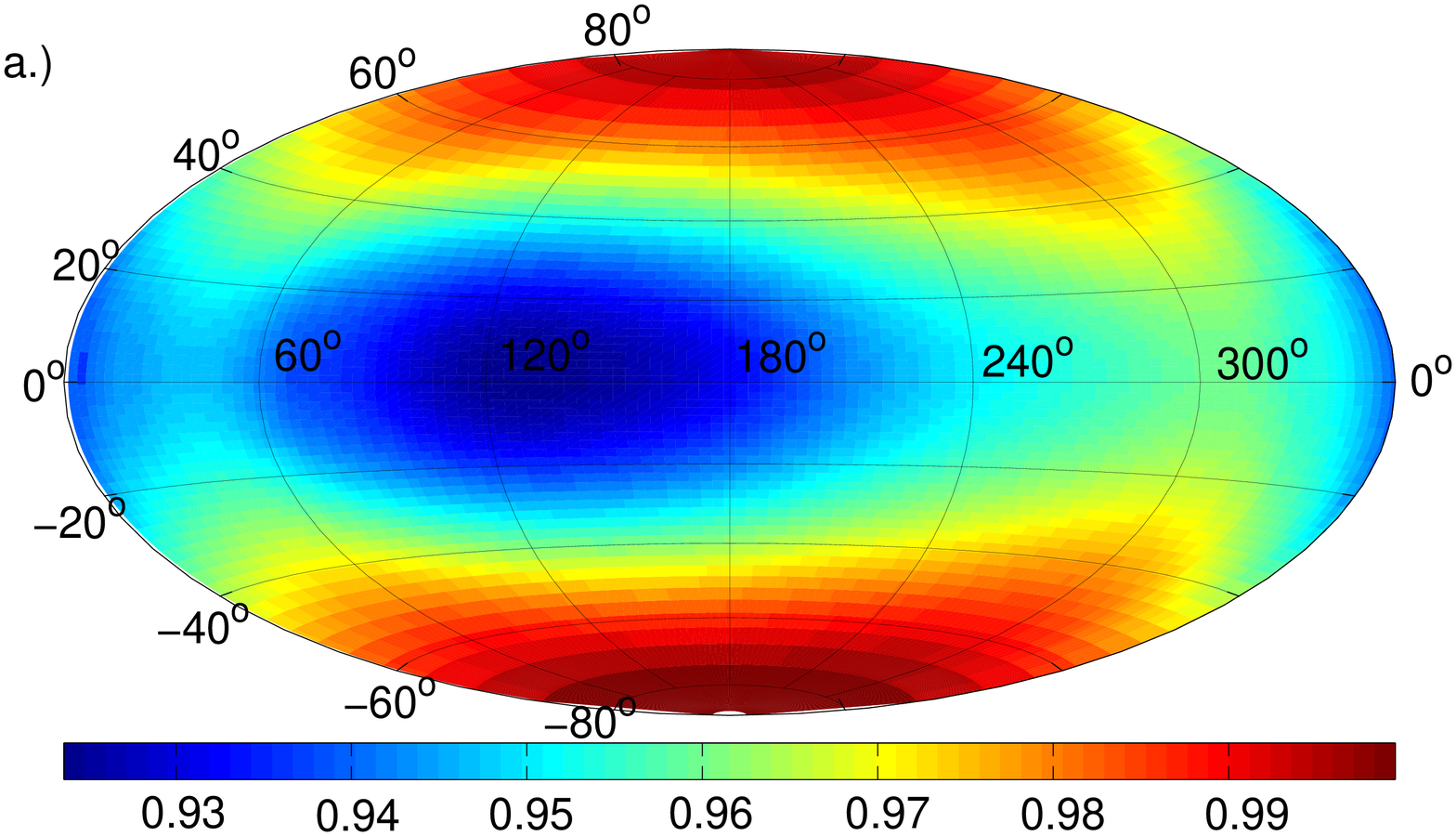}
\includegraphics*[trim = 25mm 0mm 25mm 0mm, clip, scale=.40, angle=0]{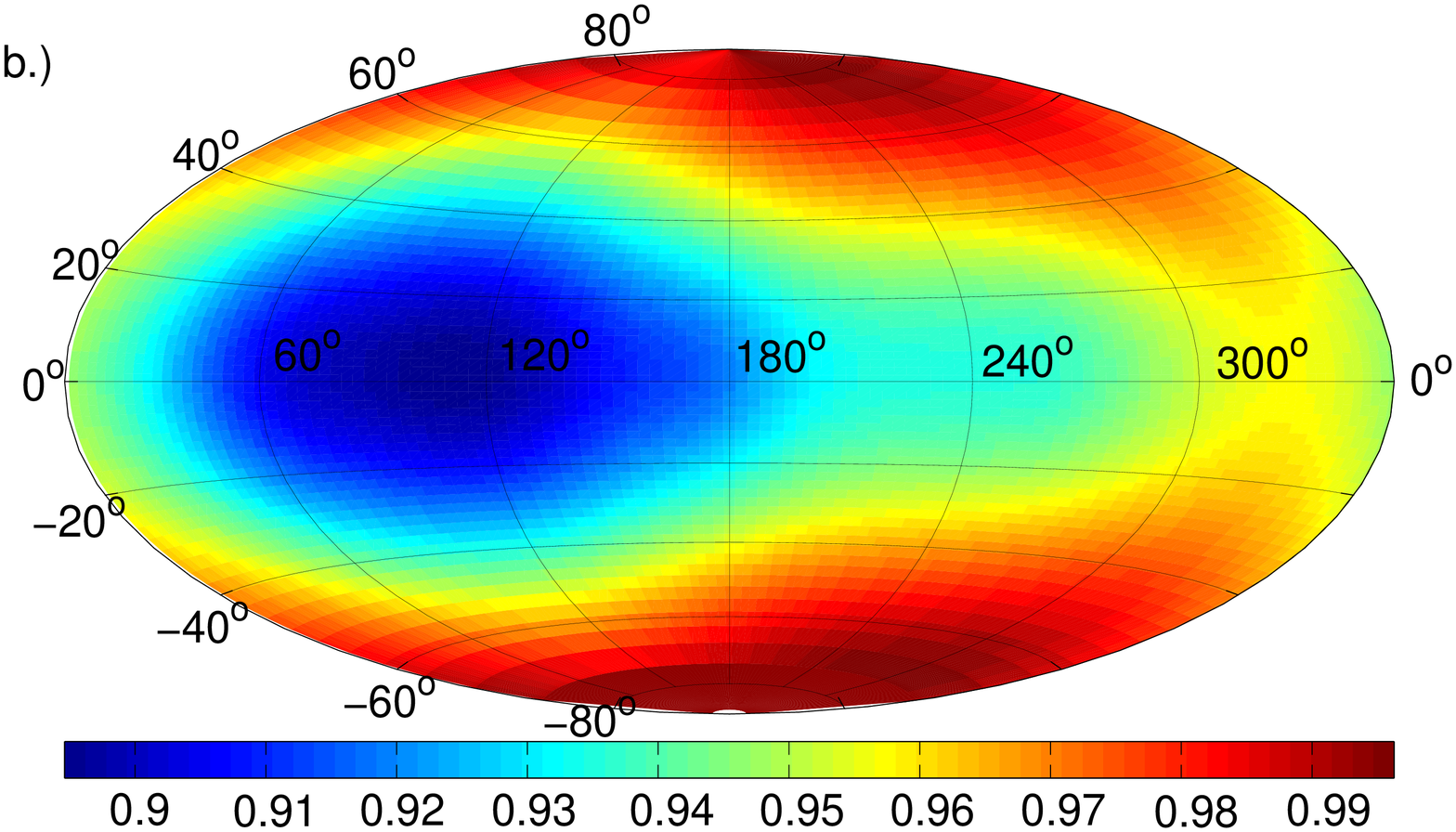}
\caption{Sky map of matches between the 1:1 (equal mass) and 4:1 waveforms for Advanced LIGO (a.) and LISA (b.).  
Throughout the text, $\phi$ is the azimuthal coordinate and $\theta$ is the polar coordinate.  The specific case shown 
corresponds to a mass of $M=100 \MSun$ for Advanced LIGO, and $(1+z)M=3\times 10^6 \MSun$ for LISA 
(the luminosity distance is irrelevant for this calculation), but the result will be qualitatively similar for any mass
$M\, \geq \,100 \MSun$ for Advanced LIGO and $M\, \geq\, 2\times 10^6 \MSun$ for LISA,
due to the constancy of the detector response and the similar spectral content of the noise over the band
of the signal for those cases, \ie cases where
the ``whitening'' procedure (see the Appendix) emphasizes the merger.
The sky location corresponds to the position on the sky of the 1:1 waveform, with the 4:1 waveform being rotated in the $\phi$ direction
to maximize the match.  Therefore, for a fixed $\theta$, the maximum in $\phi$ will correspond to the ``best'' match, and the minimum to the ``minimax'' match.}
\label{fig:skymap}
\end{center}
\end{figure*}

To further investigate the implications of the apparent degeneracy in mass ratio
for moderate-mass-ratio mergers, we calculate the match between the full 1:1
and 4:1 waveforms, including all available harmonics, as a function of the source orientation.
In Figure \ref{fig:skymap}, we show as an example a comparison with redshifted mass $(1+z)M=3\times 10^6 \MSun$ for LISA and $M=100 \MSun$
for Advanced LIGO, although the result will be qualitatively similar for any masses where the merger is emphasized 
relative to the inspiral in the whitened waveform (see the Appendix and Figure \ref{fig:white})
for a given detector (\ie any case where the signal merges at or below the peak sensitivity of the detector).  Figure \ref{fig:skymap} 
shows a sky map of the matches for Advanced LIGO and for LISA, where the match at
each point corresponds to the equal-mass waveform calculated at that point on the
source sky.  The match is maximized over the orientation of the 4:1 waveform in the $\phi$ direction
at a given $\theta$, where $\phi$ is the azimuthal coordinate and $\theta$ is the polar coordinate. 
The maximum at a given $\theta$ then corresponds to the ``best'' match, and the minimum to the ``minimax'' match
\cite{Damour:1997ub}.  
The azimuthal sky position is degenerate with the orbital phase, so that the maximization
procedure is identical to finding the maximum value in the azimuthal direction for a particular
inclination. We do not maximize over the inclination, since the spin-weighted spherical
harmonics are a more complicated function of polar angle.  In the sky maps, it is clear that the maximization over polar angle would occur
at the poles, where the quadrupole modes are most dominant.  This is consistent with our 
previous results showing the striking similarity of the quadrupole radiation across modest mass ratios.  
We observe the expected ``north/south'' symmetry, since all nonspinning binaries evolve in a fixed plane.  
The azimuthal asymmetry is greatest in the orbital plane, where the fractional luminosity of the higher 
harmonics relative to the dominant quadrupole modes is greatest.  We note that the sky map would be uniform 
for single mode matches, so the structure in Figure \ref{fig:skymap} is the result of the harmonic content, 
and therefore requires the formalism contained in the Appendix in order to maximize quasi-analytically.
The average match over the sky of the source for the cases in Figure \ref{fig:skymap} is 0.96 for Advanced LIGO, and 0.95 for LISA.  This
means that the 1:1 waveform can be considered an effective template (in the sense of \cite{Damour:1997ub}) for
typical Advanced LIGO mass ratios for a large fraction
of source orientations.

\section{Conclusions}
\label{sec:conc}

We have applied a model for nonspinning late inspiral-merger-ringdown waveforms to answer questions regarding the implications
of including the merger phase in data analysis efforts.  We have verified that, while the merger contributes a smaller fraction
of the total SNR as we deviate from the equal-mass case, it still dominates for moderate mass ratios, providing nearly
the entirety of the detectable signal for 
ground-based observations of IMBH systems.
In addition, we have studied the commonality
previously observed in the phase evolution of the merger waveform for moderate mass ratios.  While this commonality
bodes well for detection, since the equal-mass merger waveform alone would do well as a search filter for 
all moderate mass ratios, this has negative implications for signal characterization.  Indeed, by calculating
the ``match'' as a function of location on the sky of the source, we have
demonstrated that the equal-mass waveform can be considered an effective template for detecting other moderate
mass ratio signals for a 
wide range of source orientations for both Advanced LIGO and LISA.

\acknowledgments

We thank Jim van Meter and Joan Centrella for useful discussions.
STM was supported by an appointment to the NASA Postdoctoral Program at NASA Goddard Space Flight 
Center, administered by Oak Ridge Associated
Universities.

\begin{appendix}

\section*{Appendix: Generalized Phase Maximization}

We are interested in generalizing the procedure presented in \cite{Damour:1997ub} for maximizing
the match \eqref{eqn:matchdef} with respect to the initial
orbital phase constants between a target or exact (label X) waveform and a template or approximate
(label A) waveform.  Specifically, whereas the previous method is restricted in its validity to
radiation that is quadrupole--only, we wish to derive the general method for maximizing the match
for arbitrary harmonic content.  Wherever possible, we preserve the original notation from
\cite{Damour:1997ub}.

For the exact and approximate waveforms, we can represent the measured strain waveform as
\begin{widetext}
\bea
h^{A,X} (\cdots) &=& F_{+} h_{+} + F_{\times} h_{\times} = \Re\left[ F e^{\imagi \kappa} h \right] \nonumber \\
  &\equiv& \Re\left[F e^{\imagi \kappa} \sum_{\lm}\, ^{-2}Y_{\lm} (\theta,\phi)\,h_{\lm}^{A,X} (t^{A,X} - t_c^{A,X}; \varphi^{A,X}(t))\right] \nonumber \\
  & = & \Re\left[\sum_{\lm}\,F|^{-2}Y_{\lm}||h_{\lm}^{A,X}|\, e^{\imagi m\phi} e^{-\imagi m\varphi^{A,X}(t)}\right] \nonumber \\
  & = & \sum_{\lm} F |^{-2}Y_{\lm}||h_{\lm}^{A,X}|\,\left[\cos (m\phi)\,\cos (m\varphi^{A,X})+\sin (m\phi)\,\sin (m\varphi^{A,X})\right] \nonumber \\
  &\equiv& \sum_{\lm} |^{-2}Y_{\lm}|\left[ \lambda_{1m}^{A,X} h_{1\lm}^{A,X} + \lambda_{2m}^{A,X} h_{2\lm}^{A,X} \right],
\label{eqn:A_strain}
\eea
\end{widetext}
where $\lambda^{A,X}_{1m} \equiv \cos (m\varphi^{A,X})$, $\lambda^{A,X}_{2m} \equiv \sin (m\varphi^{A,X})$,
$h_{1\lm} \equiv F |h_{\lm}| \cos (m\phi)$, $h_{2\lm} \equiv F |h_{\lm}| \sin (m\phi)$, $F e^{\imagi \kappa} \equiv F_{+}+\imagi F_{\times}$ is
the complex beam pattern function,
and $\theta$ and $\phi$ describe the angular position on the source's sky (with $\kappa$ absorbed into the definition of $\phi$).  Since $\phi$ can be absorbed into $\varphi$, the following procedure
maximizes over the relative azimuthal orientation as well as the orbital phase.
For this analysis, we assume a common source polar angle $\theta$ for the exact and approximate
waveforms, although the procedure could be further generalized to allow
maximization/minimization over all relevant angles. 

As in
\cite{Damour:1997ub}, we wish to find the ``best'' and ``minimax'' match.  We therefore wish
to form an appropriate basis in which we can decompose the exact and approximate waveforms
separately, and subsequently find the projection of the resulting approximate ``vector'' on
the exact ``vector''. Conceptually, in \cite{Damour:1997ub} the procedure amounted to finding
the ellipse resulting from projecting the circle that the approximate waveform makes in its
2-plane onto  the 2-plane formed from the decomposition of the exact waveform, where the
2-planes are the spaces spanned by the orthonormal bases constructed using the exact and the approximate waveforms.
In our analysis, we extend this concept to finding the minimum and
maximum radius resulting from finding the sum of projections of approximate circles for the
available modes on the exact planes corresponding to those same modes. We could alternatively
include in the sum the cross contributions from particular approximate circles for a given mode
on the planes of all available exact modes. However, while such a result would be more directly
related to the SNR achievable by using the approximate waveform as a template, the inclusion of
cross-mode contributions would be unphysical and less useful as a gauge for potential
parameter estimation.  We therefore include only like-mode contributions, although the following
derivation can be trivially altered to include all cross-mode contributions, and the final
result will be the same in all but the most exotic cases.

\begin{figure*}
\begin{center}
\includegraphics*[scale=.25, angle=0]{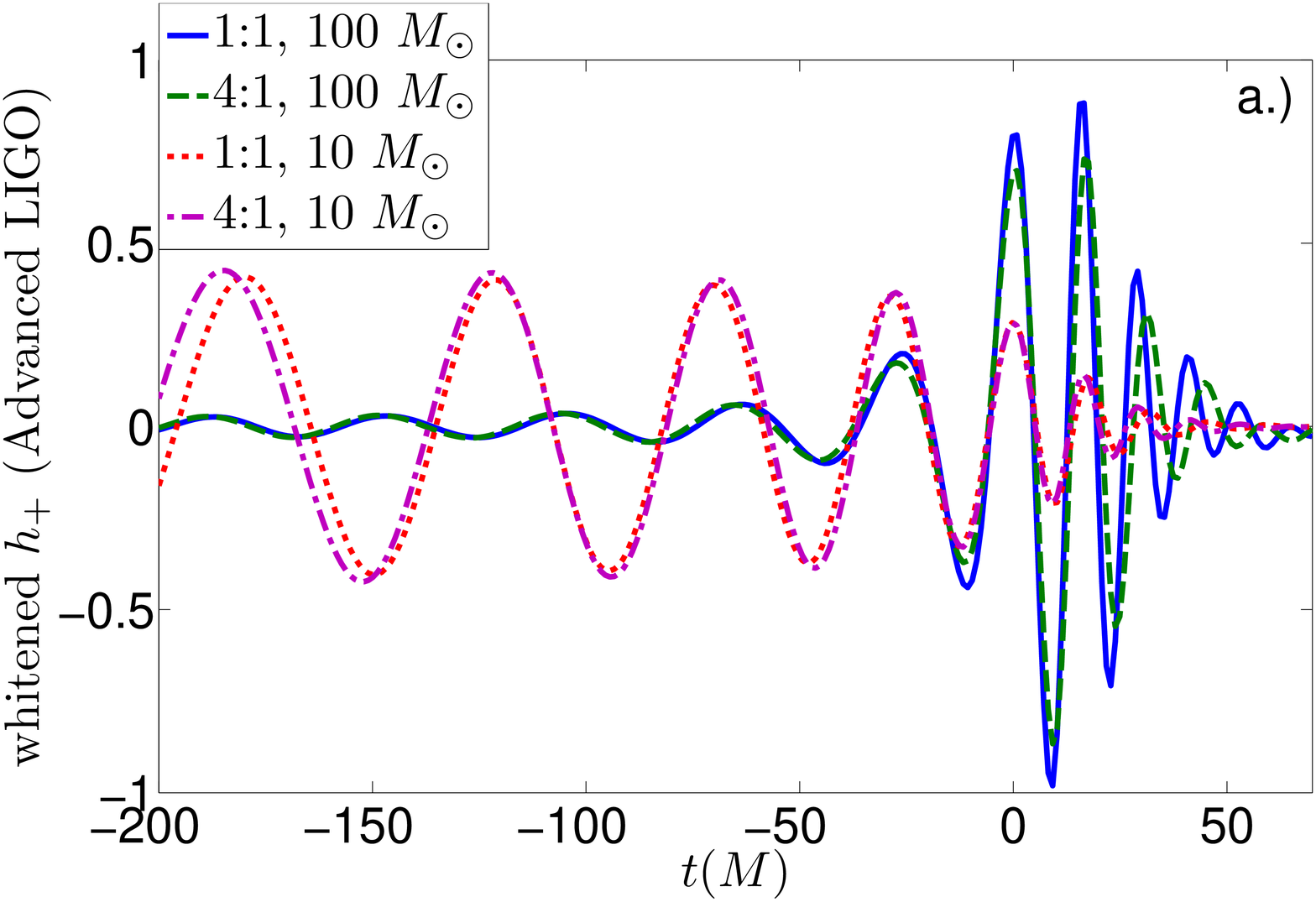}
\includegraphics*[scale=.25, angle=0]{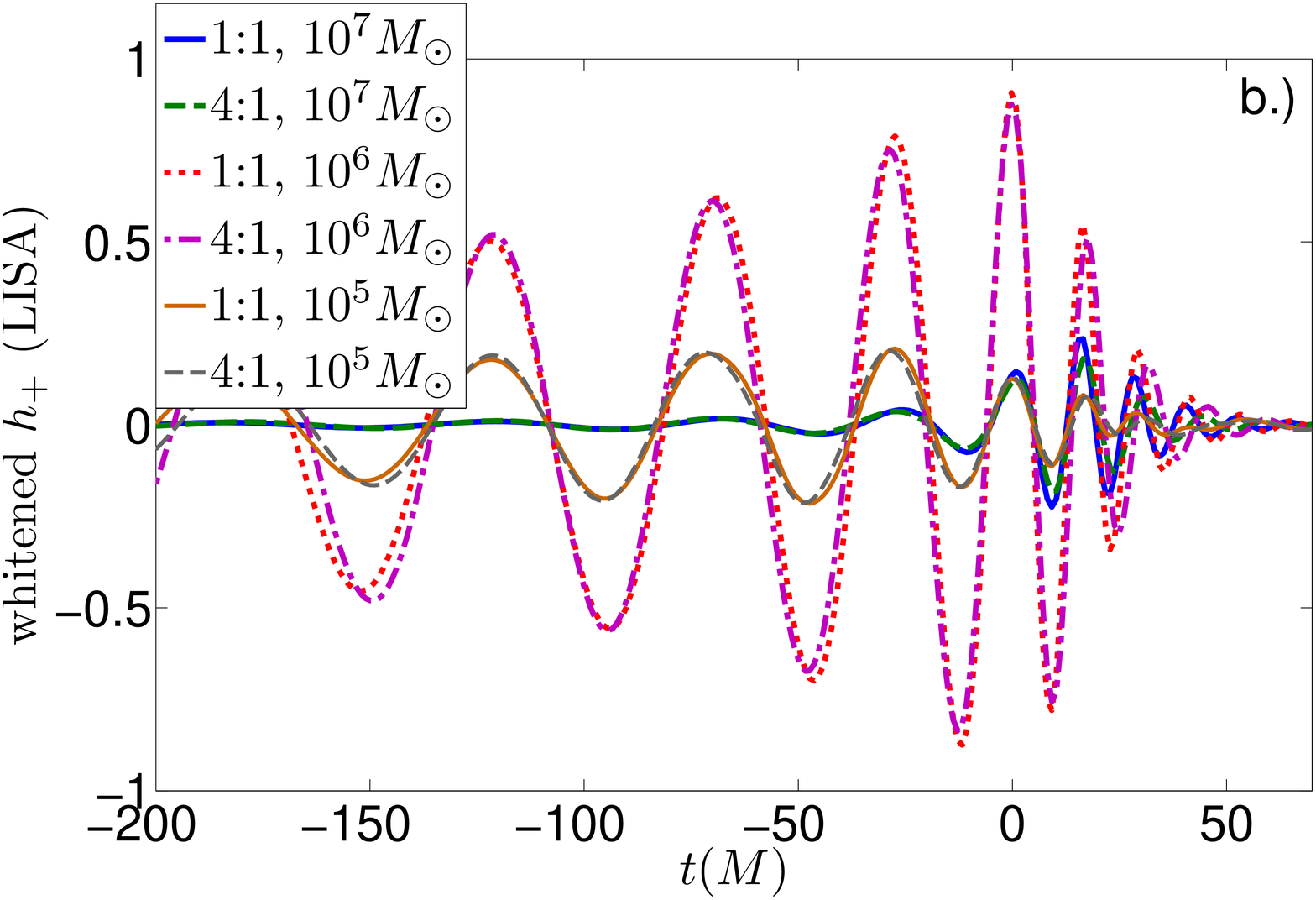}
\caption{Examples of ``whitened'' waveforms \cite{Damour:2000gg} that we use to form a basis
for calculating the match.  Examples for Advanced LIGO are shown in the left panel (a.), and examples
for LISA are shown in the right panel (b.).  The ordinate values are arbitrarily scaled.}
\label{fig:white}
\end{center}
\end{figure*}

To form the desired bases, we first construct a pair of ``whitened'' vectors \cite{Damour:2000gg},
as shown in Figure \ref{fig:white}, in both the approximate and exact planes, to account for the presence
of noise, the detector response to noise, and the detector response to the raw signal
$h_{n\lm}^{A,X}$ (where $n$ is 1 or 2),
\beq
h_{n\lm}^{A,X'} = \int\limits_{-\infty}^{+\infty} df\, \frac{\tilde{h}_{n\lm}^{A,X}}
{\sqrt{S_n}}\,e^{-\imagi 2\pi f t},
\label{eqn:A_hwhite}
\eeq
where ``$\tilde{h}$'' denotes the Fourier transform of $h$.
With these whitened vectors, the noise-weighted inner product \eqref{eqn:dotprod}
can be easily calculated in the time domain:
\beq
\langle h_1 | h_2 \rangle \equiv \int_0^{\infty} dt \, h_1'^*(t)\,h_2'(t).
\label{eqn:A_timeprod}
\eeq

Instead of attempting to construct a single orthonormal basis, we generate
an orthogonal (not normal) basis for each $\lm$ mode, with the normalization chosen
so that the sum over modes is normalized, i.e.
\bea
e_{1\lm}^{A,X} & = & \frac{h_{1\lm}^{A,X'}}{\sqrt{\sum_{\lm} \langle h_{1\lm}^{A,X'}|h_{1\lm}^{A,X'} \rangle}}, \nonumber \\
h_{2\lm}^{A,X''} & = & h_{2\lm}^{A,X'} - \langle h_{2\lm}^{A,X'} | e_{1\lm}^{A,X}\rangle e_{1\lm}^{A,X} \nonumber \\
e_{2\lm}^{A,X} & = & \frac{h_{2\lm}^{A,X''}}{\sqrt{\sum_{\lm} \langle h_{2\lm}^{A,X''}|h_{2\lm}^{A,X''} \rangle}}.
\label{eqn:A_orthogbasis}
\eea 
This expression yields an appropriate normalization over all modes, since 
$e_{n}^{A,X} \equiv \sum_{\lm} e_{n\lm}^{A,X}$ is normalized
by construction, with the individual $\lm$ modes being appropriately weighted by their
relative barycentric power and the response of the detector.  Eq.~\eqref{eqn:A_orthogbasis} is therefore a set of
orthogonal bases which are all constrained by the total signal power, and
by their common dependence on the orbital phases of the exact and approximate waveforms.  We therefore retain the
original two degrees of freedom as in \cite{Damour:1997ub}.

If we focus only on like-mode contributions to the match, we can construct the projection operator, 
$P_X(e_{\alpha}^A)$, and the resulting projection $p_{\alpha}$, of a vector $e_{\alpha}^A$ onto
the $X$-plane,
\beq
p_{\alpha}=P_X(e_{\alpha}^A)\equiv \sum_{n\lm} \frac{\langle e_{\alpha\lm}^A | e_{n\lm}^X \rangle}{\langle e_{n\lm}^X | e_{n\lm}^X \rangle} e_{n\lm}^X \, ,
\label{eqn:A_projdef}
\eeq
where $\displaystyle\sum_{n\lm}$ is shorthand for $\displaystyle\sum_{n=1}^2\sum_{\ell=2}^{\ell_{max}}
\sum_{m=-\ell}^{\ell}$, and $e_{\alpha\lm}^{A,X}$ is defined as
\beq
e_{\alpha\lm}^{A,X} \equiv \cos(m\alpha)\,e_{1\lm}^{A,X} + \sin(m\alpha)\,e_{2\lm}^{A,X} \, ,
\label{eqn:A_ealmdef}
\eeq
where $\alpha$ is an arbitrary initial orbital angle for
the approximate waveform.
%
Substituting for $e_{\alpha\lm}^{A}$ from \eqref{eqn:A_ealmdef} into
\eqref{eqn:A_projdef} yields
\beq
p_{\alpha} = \sum_{\lm} \left[ \cos (m\alpha)\ p_{1\lm} + \sin (m\alpha)\ p_{2\lm} \right] \, , 
\label{eqn:A_projlm}
\eeq
where
\beq
p_{n\lm} \equiv  P_X (e_{n\lm}^A) = \sum_{k=1}^2 \frac{\langle e_{n\lm}^A | e_{k\lm}^X \rangle}{\langle e_{k\lm}^X | e_{k\lm}^X \rangle} \, e_{k\lm}^X \, .
\label{eqn:A_pnlmdef}
\eeq
If we again focus only on like-mode contributions for simplicity, then $|p_{\alpha}|$ can be
expressed in a form which represents, geometrically, a sum of ellipses, given by
\bea
| p_{\alpha} |^2 = \sum_{\lm} & \left[ A_{\lm} \, \cos^2 (m\alpha) + B_{\lm} \, \sin^2 (m\alpha) \right. \nonumber \\
& \left. + 2 \, C_{\lm} \, \cos (m\alpha) \, \sin (m\alpha) \right] \, ,
\label{eqn:A_projsquared}
\eea
where
\bea
A_{\lm} & \equiv & | p_{1\lm} |^2 \nonumber \\
        & = & \frac{\langle e_{1\lm}^A | e_{1\lm}^X \rangle^2}{\langle e_{1\lm}^X | e_{1\lm}^X \rangle} + \frac{\langle e_{1\lm}^A | e_{2\lm}^X \rangle^2}{\langle e_{2\lm}^X | e_{2\lm}^X \rangle} \, , \nonumber \\
B_{\lm} & \equiv & | p_{2\lm} |^2 \nonumber \\
        & = & \frac{\langle e_{2\lm}^A | e_{1\lm}^X \rangle^2}{\langle e_{1\lm}^X | e_{1\lm}^X \rangle} + \frac{\langle e_{2\lm}^A | e_{2\lm}^X \rangle^2}{\langle e_{2\lm}^X | e_{2\lm}^X \rangle} \, , \nonumber \\
C_{\lm} & \equiv & \langle p_{1\lm} | p_{2\lm} \rangle \nonumber \\
        & = & \frac{\langle e_{1\lm}^A | e_{1\lm}^X \rangle\, \langle e_{2\lm}^A | e_{1\lm}^X \rangle}{\langle e_{1\lm}^X | e_{1\lm}^X \rangle} \nonumber \\
        &   & + \frac{\langle e_{1\lm}^A | e_{2\lm}^X \rangle\, \langle e_{2\lm}^A | e_{2\lm}^X \rangle}{\langle e_{2\lm}^X | e_{2\lm}^X \rangle} \, . \label{eqn:A_ABCdefs}
\eea

While \eqref{eqn:A_projsquared} is trivial to maximize or minimize analytically for the case of a single
mode as in \cite{Damour:1997ub}, the case of multiple modes generally requires a numerical solution.
However, if we assume a single mode (or mode pair) is significantly larger 
than any other mode, then
we can specify an approximate solution for the value of $\alpha$ that yields the ``best'' match. 
Generally, the condition for
extremizing \eqref{eqn:A_projsquared} is given by
\bea
\sum_{\lm}
\left[ m \, (A_{\lm}-B_{\lm}) \, \sin (2m\alpha) \right. && \nonumber \\
\left. - 2\,m \, C_{\lm} \, \cos (2m\alpha) \right] && = 0 \, .
\label{eqn:A_extremecond}
\eea
We can then apply the aforementioned assumption that a single mode pair dominates.  In geometric terms,
this means that we assume that the semi-major axis for the dominant mode(s) in Eq.~\eqref{eqn:A_projsquared} is larger than
the quadrature sum of the semi-major axes of all other modes.  In this case, the largest value for Eq.~\eqref{eqn:A_projsquared}
will occur very near the $\alpha$ that maximizes the dominant mode(s), with the other modes providing at most
a small perturbation.
This condition can be
expressed as
\bea
\frac{A_{\lmopt}+B_{\lmopt}}{2} &+& \sqrt{\left(\frac{A_{\lmopt}-B_{\lmopt}}{2}\right)^2 + C_{\lmopt}^2 }\, 
\geq\,  \nonumber \\
{\sum_{\lm \neq \lmopt }}\, \frac{A_{\lm}+B_{\lm}}{2} &+& \sqrt{\left(\frac{A_{\lm}-B_{\lm}}{2}\right)^2 
+ C_{\lm}^2 } \, ,
\label{eqn:A_dominantcond}
\eea
where $\lmopt$ corresponds to the dominant mode(s), and we only include the larger roots of Eq.~\eqref{eqn:A_extremecond}
corresponding to the ``best'' match.  For all cases in this paper,
$\lopt = |\mopt| = 2$, with $|p_{\alpha\lmopt}|=|p_{\alpha\lopt(-\mopt)}|$ by symmetry,
so that the condition for \eqref{eqn:A_dominantcond} in this case will be $\lm \neq \lopt|\mopt|$.
Finally, we can calculate the condition on $\alpha$ for maximizing the match under these assumptions:
\bea
\alpha_{\rm best} \approx\frac{1}{2\mopt} \, \cos^{-1} \left( \frac{(A_{\lmopt}-B_{\lmopt})}{\sqrt{(A_{\lmopt}-B_{\lmopt})^2 + 4C_{\lmopt}^2}} \right).
\label{eqn:A_dominantbesta}
\eea

We reiterate that Eq.~\eqref{eqn:A_dominantbesta} is not valid if subdominant modes contain comparable power to the dominant
mode or mode pair, and a similar method cannot be used to find the ``minimax'' $\alpha$.  In these cases,
Eq.~\eqref{eqn:A_extremecond} can only be solved numerically.  Even if a numerical solution is required, this method 
is still more efficient than a brute-force maximization over $\varphi^{A}$ and $\varphi^{X}$, as it makes it a 
one-dimensional search over $\alpha$.  Since $A_{\lm}$, $B_{\lm}$, and $C_{\lm}$ are all less than unity, 
the error in the match will be of the same order as the sampling interval in $\alpha$ over
the range $(0,\,2\pi]$, assuming modes with very large $m$ are negligible.  For instance,
in this work we only include $\ell \leq 4$ modes, with $m=\pm 4$ the largest relevant $m$ mode.  Therefore,
in order to calculate the match to three significant digits, we take $10^4$ samples of $\alpha$ and record
the global maximum and minimum, corresponding to the ``best'' and ``minimax'' matches, respectively.
One could implement a more clever algorithm, such as Brent's method, if the required accuracy for the match
makes the sampling procedure too computationally expensive.  We have verified that the minimum and
maximum from Eq.~\eqref{eqn:A_extremecond} agrees with the minimum and
maximum found using a Nelder-Mead simplex over the two-dimensional $\varphi^{A}$-$\varphi^{X}$ space.

We note that, even if we were to include the cross-mode contributions in our derivation, the result would remain
a sum of a set, albeit a much larger set, of ellipses.
It should be noted that the exclusion of cross-mode content makes this representation of generalized 
matches less closely related to a matched-filtered SNR calculation or even, potentially, to
the maximum likelihood estimator, in that it does not account for circumstances where the
maximum likelihood occurs at an incorrect value for the parameters.  In such cases, this
method will instead ignore all maxima except the local maximum determined from the largest
match of like-modes.
In most cases, the contribution of cross-mode terms to any match calculation will be negligible
compared to like-mode contributions,
so this local maximum will also be the global maximum.
In that case, the optimized parameter choice found from this procedure will be consistent with the
maximum likelihood value, and the resulting match will be a true representation of the fraction
of recoverable SNR from matched filtering. 

\end{appendix}


\begin{thebibliography}{32}
\expandafter\ifx\csname natexlab\endcsname\relax\def\natexlab#1{#1}\fi
\expandafter\ifx\csname bibnamefont\endcsname\relax
  \def\bibnamefont#1{#1}\fi
\expandafter\ifx\csname bibfnamefont\endcsname\relax
  \def\bibfnamefont#1{#1}\fi
\expandafter\ifx\csname citenamefont\endcsname\relax
  \def\citenamefont#1{#1}\fi
\expandafter\ifx\csname url\endcsname\relax
  \def\url#1{\texttt{#1}}\fi
\expandafter\ifx\csname urlprefix\endcsname\relax\def\urlprefix{URL }\fi
\providecommand{\bibinfo}[2]{#2}
\providecommand{\eprint}[2][]{\url{#2}}

\bibitem[{\citenamefont{Flanagan and Hughes}(1998)}]{Flanagan:1997sx}
\bibinfo{author}{\bibfnamefont{E.~E.} \bibnamefont{Flanagan}} \bibnamefont{and}
  \bibinfo{author}{\bibfnamefont{S.~A.} \bibnamefont{Hughes}},
  \bibinfo{journal}{Phys. Rev. D} \textbf{\bibinfo{volume}{57}},
  \bibinfo{pages}{4535} (\bibinfo{year}{1998}), \eprint{{arXiv}:gr-qc/9701039}.

\bibitem[{\citenamefont{Baker et~al.}(2007)\citenamefont{Baker, McWilliams, van
  Meter, Centrella, Choi, Kelly, and Koppitz}}]{Baker:2006kr}
\bibinfo{author}{\bibfnamefont{J.~G.} \bibnamefont{Baker}},
  \bibinfo{author}{\bibfnamefont{S.~T.} \bibnamefont{McWilliams}},
  \bibinfo{author}{\bibfnamefont{J.~R.} \bibnamefont{van Meter}},
  \bibinfo{author}{\bibfnamefont{J.~M.} \bibnamefont{Centrella}},
  \bibinfo{author}{\bibfnamefont{D.-I.} \bibnamefont{Choi}},
  \bibinfo{author}{\bibfnamefont{B.~J.} \bibnamefont{Kelly}}, \bibnamefont{and}
  \bibinfo{author}{\bibfnamefont{M.}~\bibnamefont{Koppitz}},
  \bibinfo{journal}{Phys. Rev. D} \textbf{\bibinfo{volume}{75}},
  \bibinfo{pages}{124024} (\bibinfo{year}{2007}),
  \eprint{{arXiv}:gr-qc/0612117}.

\bibitem[{\citenamefont{Hinder et~al.}(2008)\citenamefont{Hinder, Vaishnav,
  Herrmann, Shoemaker, and Laguna}}]{Hinder:2007qu}
\bibinfo{author}{\bibfnamefont{I.}~\bibnamefont{Hinder}},
  \bibinfo{author}{\bibfnamefont{B.}~\bibnamefont{Vaishnav}},
  \bibinfo{author}{\bibfnamefont{F.}~\bibnamefont{Herrmann}},
  \bibinfo{author}{\bibfnamefont{D.~M.} \bibnamefont{Shoemaker}},
  \bibnamefont{and} \bibinfo{author}{\bibfnamefont{P.}~\bibnamefont{Laguna}},
  \bibinfo{journal}{Phys. Rev. D} \textbf{\bibinfo{volume}{77}},
  \bibinfo{pages}{081502(R)} (\bibinfo{year}{2008}), \eprint{{arXiv}:0710.5167
  [gr-qc]}.

\bibitem[{\citenamefont{Damour et~al.}(2008)\citenamefont{Damour, Nagar,
  Dorband, Pollney, and Rezzolla}}]{Damour:2007vq}
\bibinfo{author}{\bibfnamefont{T.}~\bibnamefont{Damour}},
  \bibinfo{author}{\bibfnamefont{A.}~\bibnamefont{Nagar}},
  \bibinfo{author}{\bibfnamefont{E.~N.} \bibnamefont{Dorband}},
  \bibinfo{author}{\bibfnamefont{D.}~\bibnamefont{Pollney}}, \bibnamefont{and}
  \bibinfo{author}{\bibfnamefont{L.}~\bibnamefont{Rezzolla}},
  \bibinfo{journal}{Phys. Rev. D} \textbf{\bibinfo{volume}{77}},
  \bibinfo{pages}{084017} (\bibinfo{year}{2008}), \eprint{{arXiv}:0712.3003
  [gr-qc]}.

\bibitem[{\citenamefont{Hannam et~al.}(2007)\citenamefont{Hannam, Husa,
  Sperhake, Br\"{u}gmann, and Gonzalez}}]{Hannam:2007ik}
\bibinfo{author}{\bibfnamefont{M.~D.} \bibnamefont{Hannam}},
  \bibinfo{author}{\bibfnamefont{S.}~\bibnamefont{Husa}},
  \bibinfo{author}{\bibfnamefont{U.}~\bibnamefont{Sperhake}},
  \bibinfo{author}{\bibfnamefont{B.}~\bibnamefont{Br\"{u}gmann}},
  \bibnamefont{and} \bibinfo{author}{\bibfnamefont{J.~A.}
  \bibnamefont{Gonzalez}}, \bibinfo{journal}{Phys. Rev. D}
  \textbf{\bibinfo{volume}{77}}, \bibinfo{pages}{044020}
  (\bibinfo{year}{2007}), \eprint{{arXiv}:0706.1305 [gr-qc]}.

\bibitem[{\citenamefont{Hannam et~al.}(2008)\citenamefont{Hannam, Husa,
  Br\"{u}gmann, and Gopakumar}}]{Hannam:2007wf}
\bibinfo{author}{\bibfnamefont{M.~D.} \bibnamefont{Hannam}},
  \bibinfo{author}{\bibfnamefont{S.}~\bibnamefont{Husa}},
  \bibinfo{author}{\bibfnamefont{B.}~\bibnamefont{Br\"{u}gmann}},
  \bibnamefont{and}
  \bibinfo{author}{\bibfnamefont{A.}~\bibnamefont{Gopakumar}},
  \bibinfo{journal}{Phys. Rev. D} \textbf{\bibinfo{volume}{78}},
  \bibinfo{pages}{104007} (\bibinfo{year}{2008}), \eprint{{arXiv}:0712.3787
  [gr-qc]}.

\bibitem[{\citenamefont{Gonzalez et~al.}(2009)\citenamefont{Gonzalez, Sperhake,
  and Br\"ugmann}}]{Gonzalez:2008bi}
\bibinfo{author}{\bibfnamefont{J.~A.} \bibnamefont{Gonzalez}},
  \bibinfo{author}{\bibfnamefont{U.}~\bibnamefont{Sperhake}}, \bibnamefont{and}
  \bibinfo{author}{\bibfnamefont{B.}~\bibnamefont{Br\"ugmann}},
  \bibinfo{journal}{Phys. Rev. D} \textbf{\bibinfo{volume}{79}},
  \bibinfo{pages}{124006} (\bibinfo{year}{2009}), \eprint{{arXiv}:0811.3952
  [gr-qc]}.

\bibitem[{\citenamefont{Scheel et~al.}(2009)\citenamefont{Scheel, Boyle, Chu,
  Kidder, Matthews, and Pfeiffer}}]{Scheel:2008rj}
\bibinfo{author}{\bibfnamefont{M.~A.} \bibnamefont{Scheel}},
  \bibinfo{author}{\bibfnamefont{M.}~\bibnamefont{Boyle}},
  \bibinfo{author}{\bibfnamefont{T.}~\bibnamefont{Chu}},
  \bibinfo{author}{\bibfnamefont{L.~E.} \bibnamefont{Kidder}},
  \bibinfo{author}{\bibfnamefont{K.~D.} \bibnamefont{Matthews}},
  \bibnamefont{and} \bibinfo{author}{\bibfnamefont{H.~P.}
  \bibnamefont{Pfeiffer}}, \bibinfo{journal}{Phys. Rev. D}
  \textbf{\bibinfo{volume}{79}}, \bibinfo{pages}{024003}
  (\bibinfo{year}{2009}), \eprint{{arXiv}:0810.1767 [gr-qc]}.

\bibitem[{\citenamefont{Chu et~al.}(2009)\citenamefont{Chu, Pfeiffer, and
  Scheel}}]{Chu:2009md}
\bibinfo{author}{\bibfnamefont{T.}~\bibnamefont{Chu}},
  \bibinfo{author}{\bibfnamefont{H.~P.} \bibnamefont{Pfeiffer}},
  \bibnamefont{and} \bibinfo{author}{\bibfnamefont{M.~A.}
  \bibnamefont{Scheel}}, \bibinfo{journal}{Phys. Rev. D}
  \textbf{\bibinfo{volume}{80}}, \bibinfo{pages}{124051}
  (\bibinfo{year}{2009}), \eprint{{arXiv}:0909.1313 [gr-qc]}.

\bibitem[{\citenamefont{Ajith and Bose}(2009)}]{Ajith:2009fz}
\bibinfo{author}{\bibfnamefont{P.}~\bibnamefont{Ajith}} \bibnamefont{and}
  \bibinfo{author}{\bibfnamefont{S.}~\bibnamefont{Bose}},
  \bibinfo{journal}{Phys. Rev. D} \textbf{\bibinfo{volume}{79}},
  \bibinfo{pages}{084032} (\bibinfo{year}{2009}), \eprint{{arXiv}:0901.4936
  [gr-qc]}.

\bibitem[{\citenamefont{McWilliams}(2008)}]{McWilliams_PhD}
\bibinfo{author}{\bibfnamefont{S.~T.} \bibnamefont{McWilliams}}, Ph.D. thesis,
  \bibinfo{school}{The University of Maryland}, \bibinfo{address}{College Park,
  Maryland} (\bibinfo{year}{2008}).

\bibitem[{\citenamefont{Babak et~al.}(2008)\citenamefont{Babak, Hannam, Husa,
  and Schutz}}]{Babak:2008bu}
\bibinfo{author}{\bibfnamefont{S.}~\bibnamefont{Babak}},
  \bibinfo{author}{\bibfnamefont{M.~D.} \bibnamefont{Hannam}},
  \bibinfo{author}{\bibfnamefont{S.}~\bibnamefont{Husa}}, \bibnamefont{and}
  \bibinfo{author}{\bibfnamefont{B.~F.} \bibnamefont{Schutz}}
  (\bibinfo{year}{2008}), \bibinfo{note}{{arXiv}:0806.1591 [gr-qc]}.

\bibitem[{\citenamefont{Thorpe et~al.}(2009)\citenamefont{Thorpe, McWilliams,
  Kelly, Fahey, Arnaud, and Baker}}]{Thorpe:2008wh}
\bibinfo{author}{\bibfnamefont{J.~I.} \bibnamefont{Thorpe}},
  \bibinfo{author}{\bibfnamefont{S.~T.} \bibnamefont{McWilliams}},
  \bibinfo{author}{\bibfnamefont{B.~J.} \bibnamefont{Kelly}},
  \bibinfo{author}{\bibfnamefont{R.~P.} \bibnamefont{Fahey}},
  \bibinfo{author}{\bibfnamefont{K.}~\bibnamefont{Arnaud}}, \bibnamefont{and}
  \bibinfo{author}{\bibfnamefont{J.~G.} \bibnamefont{Baker}},
  \bibinfo{journal}{Class. Quantum Grav.} \textbf{\bibinfo{volume}{26}},
  \bibinfo{pages}{094026} (\bibinfo{year}{2009}), \bibinfo{note}{proceedings of
  the 7th International {LISA} Symposium, Barcelona, Spain, 16--20 June 2008},
  \eprint{{arXiv}:0811.0833 [astro-ph]}.

\bibitem[{\citenamefont{McWilliams et~al.}(2010)\citenamefont{McWilliams,
  Thorpe, Baker, and Kelly}}]{McWilliams:2009bg}
\bibinfo{author}{\bibfnamefont{S.~T.} \bibnamefont{McWilliams}},
  \bibinfo{author}{\bibfnamefont{J.~I.} \bibnamefont{Thorpe}},
  \bibinfo{author}{\bibfnamefont{J.~G.} \bibnamefont{Baker}}, \bibnamefont{and}
  \bibinfo{author}{\bibfnamefont{B.~J.} \bibnamefont{Kelly}},
  \bibinfo{journal}{Phys. Rev. D} \textbf{\bibinfo{volume}{81}},
  \bibinfo{pages}{064014} (\bibinfo{year}{2010}), \eprint{{arXiv}:0911.1078
  [gr-qc]}.

\bibitem[{\citenamefont{Baker et~al.}(2008)\citenamefont{Baker, Boggs,
  Centrella, Kelly, McWilliams, and van Meter}}]{Baker:2008mj}
\bibinfo{author}{\bibfnamefont{J.~G.} \bibnamefont{Baker}},
  \bibinfo{author}{\bibfnamefont{W.~D.} \bibnamefont{Boggs}},
  \bibinfo{author}{\bibfnamefont{J.~M.} \bibnamefont{Centrella}},
  \bibinfo{author}{\bibfnamefont{B.~J.} \bibnamefont{Kelly}},
  \bibinfo{author}{\bibfnamefont{S.~T.} \bibnamefont{McWilliams}},
  \bibnamefont{and} \bibinfo{author}{\bibfnamefont{J.~R.} \bibnamefont{van
  Meter}}, \bibinfo{journal}{Phys. Rev. D} \textbf{\bibinfo{volume}{78}},
  \bibinfo{pages}{044046} (\bibinfo{year}{2008}), \eprint{{arXiv}:0805.1428
  [gr-qc]}.

\bibitem[{\citenamefont{Berti et~al.}(2008)\citenamefont{Berti, Cardoso,
  Gonzalez, Sperhake, and Br\"{u}gmann}}]{Berti:2007nw}
\bibinfo{author}{\bibfnamefont{E.}~\bibnamefont{Berti}},
  \bibinfo{author}{\bibfnamefont{V.}~\bibnamefont{Cardoso}},
  \bibinfo{author}{\bibfnamefont{J.~A.} \bibnamefont{Gonzalez}},
  \bibinfo{author}{\bibfnamefont{U.}~\bibnamefont{Sperhake}}, \bibnamefont{and}
  \bibinfo{author}{\bibfnamefont{B.}~\bibnamefont{Br\"{u}gmann}},
  \bibinfo{journal}{Class. Quantum Grav.} \textbf{\bibinfo{volume}{25}},
  \bibinfo{pages}{114035} (\bibinfo{year}{2008}), \eprint{{arXiv}:0711.1097
  [gr-qc]}.

\bibitem[{\citenamefont{Schnittman et~al.}(2008)\citenamefont{Schnittman,
  Buonanno, van Meter, Baker, Boggs, Centrella, Kelly, and
  McWilliams}}]{Schnittman:2007ij}
\bibinfo{author}{\bibfnamefont{J.~D.} \bibnamefont{Schnittman}},
  \bibinfo{author}{\bibfnamefont{A.}~\bibnamefont{Buonanno}},
  \bibinfo{author}{\bibfnamefont{J.~R.} \bibnamefont{van Meter}},
  \bibinfo{author}{\bibfnamefont{J.~G.} \bibnamefont{Baker}},
  \bibinfo{author}{\bibfnamefont{W.~D.} \bibnamefont{Boggs}},
  \bibinfo{author}{\bibfnamefont{J.~M.} \bibnamefont{Centrella}},
  \bibinfo{author}{\bibfnamefont{B.~J.} \bibnamefont{Kelly}}, \bibnamefont{and}
  \bibinfo{author}{\bibfnamefont{S.~T.} \bibnamefont{McWilliams}},
  \bibinfo{journal}{Phys. Rev. D} \textbf{\bibinfo{volume}{77}},
  \bibinfo{pages}{044031} (\bibinfo{year}{2008}), \eprint{{arXiv}:0707.0301
  [gr-qc]}.

\bibitem[{\citenamefont{Br\"{u}gmann et~al.}(2008)\citenamefont{Br\"{u}gmann,
  Gonzalez, Hannam, Husa, and Sperhake}}]{Brugmann:2007zj}
\bibinfo{author}{\bibfnamefont{B.}~\bibnamefont{Br\"{u}gmann}},
  \bibinfo{author}{\bibfnamefont{J.~A.} \bibnamefont{Gonzalez}},
  \bibinfo{author}{\bibfnamefont{M.~D.} \bibnamefont{Hannam}},
  \bibinfo{author}{\bibfnamefont{S.}~\bibnamefont{Husa}}, \bibnamefont{and}
  \bibinfo{author}{\bibfnamefont{U.}~\bibnamefont{Sperhake}},
  \bibinfo{journal}{Phys. Rev. D} \textbf{\bibinfo{volume}{77}},
  \bibinfo{pages}{124047} (\bibinfo{year}{2008}), \eprint{{arXiv}:0707.0135
  [gr-qc]}.

\bibitem[{\citenamefont{Owen}(1996)}]{Owen:1995tm}
\bibinfo{author}{\bibfnamefont{B.~J.} \bibnamefont{Owen}},
  \bibinfo{journal}{Phys. Rev. D} \textbf{\bibinfo{volume}{55}},
  \bibinfo{pages}{6749} (\bibinfo{year}{1996}), \eprint{{arXiv}:gr-qc/9511032}.

\bibitem[{\citenamefont{Shoemaker}(2006)}]{DHSCom}
\bibinfo{author}{\bibfnamefont{D.}~\bibnamefont{Shoemaker}}
  (\bibinfo{year}{2006}), \bibinfo{note}{private communication}.

\bibitem[{\citenamefont{Larson et~al.}(2000)\citenamefont{Larson, Hiscock, and
  Hellings}}]{Larson:1999we}
\bibinfo{author}{\bibfnamefont{S.~L.} \bibnamefont{Larson}},
  \bibinfo{author}{\bibfnamefont{W.~A.} \bibnamefont{Hiscock}},
  \bibnamefont{and} \bibinfo{author}{\bibfnamefont{R.~W.}
  \bibnamefont{Hellings}}, \bibinfo{journal}{Phys. Rev. D}
  \textbf{\bibinfo{volume}{62}}, \bibinfo{pages}{062001}
  (\bibinfo{year}{2000}), \eprint{{arXiv}:gr-qc/9909080}.

\bibitem[{\citenamefont{Larson}()}]{LISASenGen}
\bibinfo{author}{\bibfnamefont{S.~L.} \bibnamefont{Larson}},
  \bibinfo{note}{\url{http://www.srl.caltech.edu/~shane/sensitivity}}.

\bibitem[{\citenamefont{Merkowitz}(2006)}]{MerkowitzCom}
\bibinfo{author}{\bibfnamefont{S.~M.} \bibnamefont{Merkowitz}}, in
  \emph{\bibinfo{booktitle}{Sixth International LISA Symposium}}, edited by
  \bibinfo{editor}{\bibfnamefont{S.~M.} \bibnamefont{Merkowitz}}
  \bibnamefont{and} \bibinfo{editor}{\bibfnamefont{J.~C.} \bibnamefont{Livas}}
  (\bibinfo{publisher}{American Institute of Physics}, \bibinfo{address}{New
  York}, \bibinfo{year}{2006}), pp. \bibinfo{pages}{133--142}.

\bibitem[{\citenamefont{Berti et~al.}(2006)\citenamefont{Berti, Cardoso, and
  Will}}]{Berti:2005ys}
\bibinfo{author}{\bibfnamefont{E.}~\bibnamefont{Berti}},
  \bibinfo{author}{\bibfnamefont{V.}~\bibnamefont{Cardoso}}, \bibnamefont{and}
  \bibinfo{author}{\bibfnamefont{C.~M.} \bibnamefont{Will}},
  \bibinfo{journal}{Phys. Rev. D} \textbf{\bibinfo{volume}{73}},
  \bibinfo{pages}{064030} (\bibinfo{year}{2006}),
  \eprint{{arXiv}:gr-qc/0512160}.

\bibitem[{\citenamefont{Goldberg et~al.}(1967)\citenamefont{Goldberg,
  MacFarlane, Newman, Rohrlich, and Sudarshan}}]{Goldberg:1967}
\bibinfo{author}{\bibfnamefont{J.~N.} \bibnamefont{Goldberg}},
  \bibinfo{author}{\bibfnamefont{A.~J.} \bibnamefont{MacFarlane}},
  \bibinfo{author}{\bibfnamefont{E.~T.} \bibnamefont{Newman}},
  \bibinfo{author}{\bibfnamefont{F.}~\bibnamefont{Rohrlich}}, \bibnamefont{and}
  \bibinfo{author}{\bibfnamefont{E.~C.~G.} \bibnamefont{Sudarshan}},
  \bibinfo{journal}{J. Math. Phys.} \textbf{\bibinfo{volume}{8}},
  \bibinfo{pages}{2155} (\bibinfo{year}{1967}).

\bibitem[{\citenamefont{Buonanno et~al.}(2007)\citenamefont{Buonanno, Pan,
  Baker, Centrella, Kelly, McWilliams, and van Meter}}]{Buonanno:2007pf}
\bibinfo{author}{\bibfnamefont{A.}~\bibnamefont{Buonanno}},
  \bibinfo{author}{\bibfnamefont{Y.}~\bibnamefont{Pan}},
  \bibinfo{author}{\bibfnamefont{J.~G.} \bibnamefont{Baker}},
  \bibinfo{author}{\bibfnamefont{J.~M.} \bibnamefont{Centrella}},
  \bibinfo{author}{\bibfnamefont{B.~J.} \bibnamefont{Kelly}},
  \bibinfo{author}{\bibfnamefont{S.~T.} \bibnamefont{McWilliams}},
  \bibnamefont{and} \bibinfo{author}{\bibfnamefont{J.~R.} \bibnamefont{van
  Meter}}, \bibinfo{journal}{Phys. Rev. D} \textbf{\bibinfo{volume}{76}},
  \bibinfo{pages}{104049} (\bibinfo{year}{2007}), \eprint{{arXiv}:0706.3732
  [gr-qc]}.

\bibitem[{\citenamefont{Detweiler and Szedenits}(1979)}]{Detweiler79}
\bibinfo{author}{\bibfnamefont{S.~L.} \bibnamefont{Detweiler}}
  \bibnamefont{and}
  \bibinfo{author}{\bibfnamefont{E.}~\bibnamefont{Szedenits}},
  \bibinfo{journal}{Astrophys. J.} \textbf{\bibinfo{volume}{231}},
  \bibinfo{pages}{211} (\bibinfo{year}{1979}).

\bibitem[{\citenamefont{Damour et~al.}(2000{\natexlab{a}})\citenamefont{Damour,
  Jaranowski, and Sch\"{a}fer}}]{Damour:2000we}
\bibinfo{author}{\bibfnamefont{T.}~\bibnamefont{Damour}},
  \bibinfo{author}{\bibfnamefont{P.}~\bibnamefont{Jaranowski}},
  \bibnamefont{and}
  \bibinfo{author}{\bibfnamefont{G.}~\bibnamefont{Sch\"{a}fer}},
  \bibinfo{journal}{Phys. Rev. D} \textbf{\bibinfo{volume}{62}},
  \bibinfo{pages}{084011} (\bibinfo{year}{2000}{\natexlab{a}}),
  \eprint{{arXiv}:gr-qc/0005034}.

\bibitem[{\citenamefont{Cook}(2002)}]{Cook:2001wi}
\bibinfo{author}{\bibfnamefont{G.~B.} \bibnamefont{Cook}},
  \bibinfo{journal}{Phys. Rev. D} \textbf{\bibinfo{volume}{65}},
  \bibinfo{pages}{084003} (\bibinfo{year}{2002}),
  \eprint{{arXiv}:gr-qc/0108076}.

\bibitem[{\citenamefont{Lindblom et~al.}(2008)\citenamefont{Lindblom, Owen, and
  Brown}}]{Lindblom:2008cm}
\bibinfo{author}{\bibfnamefont{L.}~\bibnamefont{Lindblom}},
  \bibinfo{author}{\bibfnamefont{B.}~\bibnamefont{Owen}}, \bibnamefont{and}
  \bibinfo{author}{\bibfnamefont{D.~A.} \bibnamefont{Brown}},
  \bibinfo{journal}{Phys. Rev. D} \textbf{\bibinfo{volume}{78}},
  \bibinfo{pages}{124020} (\bibinfo{year}{2008}), \eprint{{arXiv}:0809.3844
  [gr-qc]}.

\bibitem[{\citenamefont{Damour et~al.}(1998)\citenamefont{Damour, Iyer, and
  Sathyaprakash}}]{Damour:1997ub}
\bibinfo{author}{\bibfnamefont{T.}~\bibnamefont{Damour}},
  \bibinfo{author}{\bibfnamefont{B.~R.} \bibnamefont{Iyer}}, \bibnamefont{and}
  \bibinfo{author}{\bibfnamefont{B.~S.} \bibnamefont{Sathyaprakash}},
  \bibinfo{journal}{Phys. Rev. D} \textbf{\bibinfo{volume}{57}},
  \bibinfo{pages}{885} (\bibinfo{year}{1998}), \eprint{{arXiv}:gr-qc/9708034}.

\bibitem[{\citenamefont{Damour et~al.}(2000{\natexlab{b}})\citenamefont{Damour,
  Iyer, and Sathyaprakash}}]{Damour:2000gg}
\bibinfo{author}{\bibfnamefont{T.}~\bibnamefont{Damour}},
  \bibinfo{author}{\bibfnamefont{B.~R.} \bibnamefont{Iyer}}, \bibnamefont{and}
  \bibinfo{author}{\bibfnamefont{B.~S.} \bibnamefont{Sathyaprakash}},
  \bibinfo{journal}{Phys. Rev. D} \textbf{\bibinfo{volume}{62}},
  \bibinfo{pages}{084036} (\bibinfo{year}{2000}{\natexlab{b}}),
  \eprint{{arXiv}:gr-qc/0001023}.

\end{thebibliography}
\end{document}